\documentclass[pra,aps,epsf,nofootinbib,twocolumn,
notitlepage,showpacs,10pt,floatfix]{revtex4-1}

\usepackage{color}
\usepackage{float}
\usepackage{hyperref}
\usepackage{graphicx}
\usepackage{subcaption}
\usepackage{amsmath}
\usepackage{amsthm}
\usepackage{mathtools}
\usepackage{bbold}

\theoremstyle{definition}

\usepackage{bm}
\newcommand{\argmax}{\operatornamewithlimits{arg\ max}}

\newcommand{\Beta}{\mathrm{B}}
\theoremstyle{definition}

\usepackage{graphicx} 
\graphicspath{ {rrImages/} } 
\usepackage{amsfonts}
\usepackage{amssymb}
\newcommand{\etal}{{\it et al.~}}
\usepackage[normalem]{ulem}
\begin{document}

\title{Empirical performance bounds for quantum approximate optimization}

\author{Phillip C. Lotshaw}
\email{lotshawpc@ornl.gov}
\affiliation{
	Quantum Computational Sciences Group\\ Oak Ridge National Laboratory\\ Oak Ridge, Tennessee 37830 USA}
	
\author{Travis S. Humble}
\email{humblets@ornl.gov}

\thanks{\\ This manuscript has been authored by UT-Battelle, LLC under Contract No. DE-AC05-00OR22725 with the U.S. Department of Energy. The United States Government retains and the publisher, by accepting the article for publication, acknowledges that the United States Government retains a non-exclusive, paid-up, irrevocable, world-wide license to publish or reproduce the published form of this manuscript, or allow others to do so, for United States Government purposes. The Department of Energy will provide public access to these results of federally sponsored research in accordance with the DOE Public Access Plan. (http://energy.gov/downloads/doe-public-access-plan).
}

\affiliation{
	Quantum Computational Sciences Group\\ Oak Ridge National Laboratory\\ Oak Ridge, Tennessee 37830 USA}

\author{Rebekah Herrman}
\email{rherrma2@utk.edu}
\affiliation{
	Department of Industrial and Systems Engineering, University of Tennessee at Knoxville\\Knoxville, Tennessee  37996-2315 USA}

\author{James Ostrowski}
\email{jostrows@utk.edu}
\affiliation{
	Department of Industrial and Systems Engineering, University of Tennessee at Knoxville\\Knoxville, Tennessee  37996-2315 USA}

\author{George Siopsis}
\email{siopsis@tennessee.edu}
\affiliation{
	Department of Physics and Astronomy, University of Tennessee at Knoxville\\Knoxville, Tennessee 37996-1200 USA}

\begin{abstract}
The quantum approximate optimization algorithm (QAOA) is a variational method for noisy, intermediate-scale quantum computers to solve combinatorial optimization problems. 
Quantifying performance bounds with respect to specific problem instances provides insight into when QAOA may be viable for solving real-world applications. 
Here, we solve every instance of  MaxCut on non-isomorphic unweighted graphs with nine or fewer vertices by numerically simulating the pure-state dynamics of QAOA.
Testing up to three layers of QAOA depth, we find that distributions of the approximation ratio narrow with increasing depth while the probability of recovering the maximum cut generally broadens.  We find QAOA exceeds the Goemans-Williamson approximation ratio bound for most graphs.
We also identify consistent patterns within the ensemble of optimized variational circuit parameters that offer highly efficient heuristics for solving MaxCut with QAOA.
The resulting data set is presented as a benchmark for establishing empirical bounds on QAOA performance that may be used to test on-going experimental realizations. 
\end{abstract}

\maketitle


\section{Introduction}
Noisy, intermediate-scale quantum (NISQ) computers may soon solve problems of practical importance with a quantum computational advantage \cite{arute2019quantum,Zhongeabe8770}.  One promising approach uses the quantum approximate optimization algorithm (QAOA) \cite{farhi2014quantum,wang2018quantum,Hadfield2018dissertation,zhou2020quantum,MaxCutRequiresHundredsQubits,Medvidovic2020QAOA54qubit,brandao2018concentration,Love2020Bounds,Shaydulin2020CaseStudy,crooks2018performance,Shaydulin2020Symmetries,Herrman2020depth,ReachabilityDeficit,Szegedy2020GraphQAOA,Pagano2020TrappedIonQAOA} or its variants \cite{wang2020xy,zhu2020adaptqaoa,Jiang2017GroverQAOA,Eidenbenz2020GroverMixers,Bartschi2020MaxkCover,LiLi2020Gibbs,Gupta2020WarmStart} to find approximate solutions to combinatorial optimization problems. By using alternating layers of a mixing operator and a cost operator, QAOA promises to prepare an approximation to the quantum state that maximizes the expectation value of the cost operator \cite{farhi2014quantum}. Moreover, the cost operator may represent an instance of unconstrained combinatorial optimization, which opens the application of QAOA to a wide variety of practical but challenging computational problems, including the well known graph problem MaxCut.
\par 
Alongside theoretical guarantees, the empirical performance of QAOA is an important open question for evaluating computational utility. The optimal number of alternating circuit layers as well as their tuning has been observed to vary with specific problem instances.  Previous efforts examining MaxCut have tested performance in terms of the approximation ratio, which quantifies the average cost function value observed relative to the optimal value. These include studies on families of problem instances represented by 2-regular graphs \cite{farhi2014quantum,wang2018quantum,Hadfield2018dissertation}, 3-regular graphs \cite{zhou2020quantum,brandao2018concentration,Medvidovic2020QAOA54qubit,MaxCutRequiresHundredsQubits,Love2020Bounds}, small samples of random graphs \cite{crooks2018performance,Shaydulin2020CaseStudy}, and small samples of graphs with various fixed symmetries \cite{Shaydulin2020Symmetries}. Notably, Crooks has shown that QAOA may exceed the performance bounds set by the Goemans-Williamson algorithm \cite{crooks2018performance}, which yields the best conventional lower bound on the approximation ratio for MaxCut. This collection of results has encouraged further study of the more general circumstances under which QAOA may provide a practical improvement in performance. 
\par
An additional measure of QAOA performance is the resources required to prepare the optimized circuit layers. By design, the alternating layers of QAOA are tuned to prepare the approximate quantum state using parameterized gates optimized with respect to the observed cost value. 
However, the effort to identify these optimized circuits becomes intractable for large numbers of parameters.  Previous studies have used optimization heuristics based on machine learning \cite{crooks2018performance,Shaydulin2020Symmetries} and various numerical optimization algorithms \cite{zhou2020quantum,Bartschi2020MaxkCover,Shaydulin2020CaseStudy,MaxCutRequiresHundredsQubits,Medvidovic2020QAOA54qubit}. These results have been limited to small sets of graphs, often with simple regular structures, and an open question is whether there are general heuristics that yield good parameters quickly and consistently.
\par
Here we use numerical simulations of pure state dynamics to quantify performance bounds of QAOA solving an exhaustive set of MaxCut instances for $n\leq9$ vertex graphs at depths $p \leq 3$. The exhaustive problem-instance set gives a thorough and systematic account of QAOA behavior on MaxCut for small graphs.  We also evaluate the effectiveness of circuit optimization using the  Broyden-Fletcher-Goldfarb-Shanno (BFGS) algorithm \cite{NumericalRecipesBFGS} by benchmarking against exact optimization software and brute force solutions for all graphs at $p=1$ and for graphs with $n \leq 6$ vertices at $p=2$.  We confirm BFGS returns optimal angles for QAOA state preparation in all these cases. Ultimately, our simulation results reveal patterns in the optimized circuits that support heuristics for more efficient parameter selection. 
\par
Our characterizations of QAOA performance for solving MaxCut address the approximation ratio and the probability of obtaining the optimal result. While the former measure estimates the average cost function value returned by QAOA relative to the actual optimal value, the latter quantifies the probability to recover the optimal cut. We compile the results for each graph instance into a data set \cite{dataset} that serves as a validated benchmark to support experimental testing of QAOA on NISQ devices as well as analyses of how specific graph structure features are correlated with performance \cite{herrman2021graphprop}. 
\par
The remainder of the presentation is organized as follows: we review QAOA in Sec.~\ref{background} and discuss our approach to variational circuit simulations in Sec.~\ref{calculations}. We present results from these simulations in Sec.~\ref{results}, including trends in the performance measures and gate parameters, before concluding with our key findings in Sec.~\ref{conclusion}.

\section{Quantum Approximate Optimization Algorithm} \label{background}

The quantum approximate optimization algorithm (QAOA) is a variational algorithm designed to find good approximate solutions to combinatorial optimization problems \cite{farhi2014quantum}.  A problem instance is specified by a cost function $C(z)$ with $z = (z_{n-1}, z_{n-2}, ..., z_0)$ and $z_j \in \{0,1\}$. QAOA recovers a candidate solution $z^*$ and the quality of this candidate can be quantified by the normalized value of the cost function
\begin{equation} \label{rz} \mathcal{R}(z^*)=\frac{C(z^*)}{C_\mathrm{max}}.
\end{equation}
\noindent Here 
\begin{equation} \label{cmax} C_\mathrm{max} = \max_z\ C(z) \end{equation} 
\noindent is the globally optimal value with optimal solution 
\begin{equation} 
\label{zmax} z_\mathrm{max} = \argmax_z 
\ C(z)
\end{equation}
\noindent and $\mathcal{R}(z_\mathrm{max})=1$ while $\mathcal{R} < 1$ otherwise.

QAOA encodes the bitstring $z$ as a quantum state with respect to the computational basis $\vert z_j \rangle \in \{\vert 0 \rangle, \vert 1 \rangle\}$ such that $\vert z \rangle = \vert z_{n-1},...,z_0\rangle$.  The cost function $C(z)$ is expressed as the operator $\hat C$ that is diagonal in the computational basis with matrix elements 
\begin{equation} 
\label{cost operator generic} \langle z \vert \hat C \vert z \rangle = C(z).
\end{equation}

The initial state of QAOA is taken as a uniform superposition of the computational basis states
\begin{equation} \label{initial state} \vert \psi_0\rangle = \frac{1}{\sqrt{2^n}} \sum_{z=0}^{2^{n}-1} \vert z \rangle \end{equation}
\noindent that is transformed by a series of $p$ unitary operations as
\begin{equation} \label{psi p} \vert \psi_p(\bm \gamma, \bm \beta)\rangle = \left(\prod_{q=1}^p \hat U(\hat B, \beta_q)\hat U(\hat C, \gamma_q)\right) \vert \psi_0\rangle \end{equation}
\noindent with variational angle parameters $\bm \gamma = (\gamma_1,...,\gamma_p)$ and $\bm \beta = (\beta_1,...,\beta_p)$.  The first unitary operator 
\begin{equation} \label{UC} \hat U(\hat C, \gamma_q) = \exp(-i \gamma_q \hat C) \end{equation}
\noindent applies a $C(z)$-dependent phase to each of the computational basis states. The second unitary 
\begin{equation} \label{UB}  \hat U(\hat B, \beta_q) = \exp(-i \beta_q \hat B), \end{equation} 
\noindent applies coupling in the computational basis as
\begin{equation} \label{mixer} \hat B = \sum_{j=0}^{n-1} \hat X_j, \end{equation}
where $\hat X_j$ is the Pauli $X$ operator on qubit $j$.  The latter transitions between the $\{\vert z \rangle\}$ depend on the $C(z)$-dependent phases from $\hat U(\hat C, \gamma_q)$ and the angle parameters $(\bm \gamma, \bm \beta)$.  QAOA selects these parameters to maximize the value $\langle \hat C \rangle$, as discussed in more detail in Sec.~\ref{optimizing intro section}.

Measurement of the state  $\vert \psi_p(\bm \gamma, \bm \beta)\rangle$  in the computational basis yields each $z_j \in \{0,1\}$ and therefore a candidate solution $z^*$. The probability to observe a specific $z$ is given by the Born rule
\begin{equation} \label{pz} P(z) = |\langle z \vert \psi_p(\bm \gamma, \bm \beta)\rangle|^2. \end{equation}
\noindent  Following measurement, the observed result $z$ is used to calculate the cost $C(z)$. Repeating the sequence of preparation and measurement approximates the distribution of $z$ given by Eq.~(\ref{pz}). The proposed solution $z^*$ may then be selected as the argument that yields the largest cost function as defined by Eq.~(\ref{rz}).  
\par
As shown by Farhi \etal \cite{farhi2014quantum}, the  probability that a measurement returns the optimal solution $z_\mathrm{max}$ converges to unity as the QAOA depth $p$ goes to infinity. Less is known about the performance of QAOA at finite $p$, where studies suggest a modest $p$ may suffice for achieving a quantum computational advantage \cite{crooks2018performance,MaxCutRequiresHundredsQubits}. 

\subsection{Gate parameter optimization} \label{optimizing intro section}
Finding optimal solutions, or good approximations, to Eq.~(\ref{rz}) requires optimizing the angle parameters that tune each QAOA layer. A standard approach to optimizing these angles is to pick an initial set of values $(\bm\gamma^{(0)},\bm\beta^{(0)})$ and then prepare and measure many copies of the state $\vert \psi_p(\bm\gamma^{(0)},\bm\beta^{(0)})\rangle$ to estimate the expectation value of the cost function operator as
\begin{equation} \label{expectation c eq} \langle \hat C(\bm \gamma, \bm \beta)\rangle = \sum_z P(z) C(z). \end{equation} 
\noindent Analytically, $P(z)$ depends on $\vert \psi_p(\bm \gamma, \bm \beta)\rangle$ through Eq.~(\ref{pz}).  

When using an optimization algorithm, such as gradient descent or BFGS \cite{NumericalRecipesBFGS}, new angles $(\bm\gamma^{(1)},\bm\beta^{(1)})$ are tested for increases in $ \langle \hat C\rangle$. After $(\bm\gamma^{(1)},\bm\beta^{(1)})$ have been selected, the process is repeated with the new angles: prepare and measure a set of states $\vert \psi_p (\bm\gamma^{(1)},\bm\beta^{(1)})\rangle$, send the results $\{z \}$ to a classical computer to calculate an estimate of $\langle \hat C(\bm\gamma^{(1)},\bm\beta^{(1)})\rangle$, then pick new angles to try with the quantum computer.
The optimization process is repeated until a set of angles is found to maximize $\langle \hat C \rangle$.  States with the optimized angles are repeatedly prepared and measured to sample $z$ and identify a solution $z^*$ that gives the best approximation. 
\par
Previous studies have found patterns in the optimized parameters that can potentially simplify gate-angle optimization. Zhou \etal \cite{zhou2020quantum} examined regular graphs and developed parameter optimization approaches for deep circuits using linear interpolation or discrete cosine and sine transformations of parameters from lower depths. Brand\~ao \etal \cite{brandao2018concentration} have argued that parameters should be transferable between graphs with sufficiently similar structure, suggesting that approaches like that of Zhou \etal may be applicable within any families of structured graphs.  Other studies suggest that parameter patterns apply more generally between varying types of graphs, as observed in small samples of random graphs \cite{crooks2018performance,Shaydulin2020CaseStudy}, or perhaps even between varying problems and QAOA-like algorithms, as in an approach to Max-$k$ Vertex Cover using a variant formulation of QAOA \cite{Bartschi2020MaxkCover}.  Together these studies suggest that parameter patterns may be generic and important in simplifying optimization for QAOA. However, it is unclear if these patterns will extend over general graphs or if they are byproducts of the structures and small samples of graphs examined so far.
\par
In addition to the angle parameters, a depth parameter $p$ defines a given state preparation circuit and instance of QAOA.  Higher depths theoretically give better performance \cite{farhi2014quantum}, but they also require more computationally intensive optimizations.  It is therefore efficient to use the smallest depth needed to obtain a desired quality of solution.  Previous studies have characterized performance for 2-regular graphs at arbitrary depths \cite{farhi2014quantum,wang2018quantum} and established performance bounds for 3-regular graphs at several low depths \cite{Love2020Bounds}.  For general graphs, an analytical expression that characterizes performance has been derived for the lowest-possible depth $p=1$ \cite{wang2018quantum,Hadfield2018dissertation} and the relationship between depth and performance under specific parameter schedules has been examined for graphs with varying symmetry \cite{Shaydulin2020Symmetries}.  However, much less is known about  QAOA performance on general graphs at depths $p> 1$ and which depths suffice for applications.

\subsection{Characterizing QAOA performance} \label{success measures}

The quality of a solution $z^*$ is characterized by the normalized cost function in Eq.~(\ref{rz}).  In QAOA, $z^*$ depends on measurements that probabilistically sample the computational basis states following Eq.~(\ref{pz}).  Therefore, we characterize QAOA performance using the quantum approximation ratio
\begin{equation} \label{avg r}  r  = \langle \mathcal{R}(z)\rangle, \end{equation}

\noindent where $\mathcal{R}(z)$ is the normalized cost and the expectation value is taken as in Eq.~(\ref{expectation c eq}).   The quantity $r$ describes the average $\mathcal{R}(z)$ with respect to measurements of $\vert \psi_p(\bm \gamma, \bm \beta)\rangle$.  An argument about concentration in QAOA claims that solutions $z^*$ with $\mathcal{R}(z^*) \geq r$ can be expected with high probability given a modest number of quantum state measurements \cite{farhi2014quantum}.  The quantum $r$ is thus also related to the classical notion of an approximation ratio as a minimum bound on performance for the optimization algorithm.

A second way to characterize QAOA success quantifies the probability to obtain the optimal solution $z_\mathrm{max}$ from Eq.~(\ref{zmax}).   In general, there can be multiple optimal solutions, which we denote as the set of $k$ optimal solutions  $\{z_\mathrm{max}^i\}_{i \in [k]}$, and the probability of optimizing the cost function is 
\begin{equation} \label{pcmax}   P(C_\mathrm{max}) = \sum_{i=1}^k P(z_\mathrm{max}^i),       \end{equation}

\noindent where $P(z_\mathrm{max}^i)$ is the probability of $z_\mathrm{max}^i$ from Eq.~(\ref{pz}). Letting $P_s$ be a desired threshold to recover the max cut, the minimal number of samples $N_s$ to observe a cut that maximizes the cost function is given by
\begin{align}
    N_s &= \frac{\log{(1-P_s)}}{\log{(1-P(C_\mathrm{max}))}},
\end{align}

\noindent since the probability to observe no maximum cut in $N_s$ samples is $P' = (1-P(C_{max}))^{N_s}$ and $P' = 1-P_s$.

\subsection{MaxCut}\label{maxcut subsection}

MaxCut partitions a graph $G$ such that the number of edges shared between the partitions is maximized. Let $G=(V,E)$ with $V = \{n-1,...,0\}$ the set of vertex labels and $E = \{ \langle j,k\rangle : j,k \in V\}$  the set of unweighted edges. For each vertex $j \in V$, a cut $z = (z_{n-1},...,z_0)$ assigns a binary label $z_j \in \{0,1\}$ that denotes the corresponding set. We express the cost function for MaxCut as a sum of pair-wise clauses
\begin{equation} \label{cost function} C(z) = \sum_{\langle j,k\rangle} C_{\langle j,k\rangle}(z). \end{equation} 
with 
\begin{equation} \label{edge cost function} 
C_{\langle j,k\rangle}(z) = z_j + z_k - 2z_jz_k  = \left\{ \begin{array}{cc} 
                1 & \hspace{1mm} z_j \neq z_k\\
                0 & \hspace{1mm} z_j = z_k \\
                \end{array} \right . \end{equation}
\noindent  An individual clause $C_{\langle j,k\rangle}(z)$ is maximized when two vertices $j,k$ connected by an edge are assigned to opposing sets, and the purpose of MaxCut is to find a cut $z_\mathrm{max}$ that maximizes Eq.~(\ref{cost function}).  There are always at least two maximum cuts since the $C_{\langle j,k\rangle}$ is invariant under $0\leftrightarrow1$ for all the bits $z_j$ in any $z$.

From  Eq.~(\ref{cost operator generic}), the cost function is cast as a sum of operators for the edges
\begin{equation} \label{cost operator} \hat C = \sum_{\langle j,k\rangle} \hat C_{\langle j,k\rangle}, \end{equation}
\noindent with edge operators
\begin{equation} \label{edge cost operator} \hat C_{\langle j,k\rangle} = \frac{1}{2} \left(\hat{\mathbb{1}} - \hat Z_j \hat Z_k\right), \end{equation}

\noindent where $\hat{\mathbb{1}}$ is the identity operator and $\hat Z_j$ is the Pauli $Z$ operator on qubit $j$. The notation $\hat Z_j \hat Z_k$ uses an implicit tensor product with the identity operator on the Hilbert space of unlisted qubits. 

\section{Simulations of QAOA} \label{calculations}

We simulate QAOA to solve MaxCut of all connected non-isomorphic graphs with $n \leq 9$ vertices \cite{GraphFiles}.  We only consider connected graphs since solutions for any disconnected graph can be constructed from separate solutions of the connected graph components. The number of graphs $N_n$ grows rapidly with the number of vertices $n$, as shown in Table \ref{graph count table}.  

\begin{table}[h]
    \centering
    \begin{tabular}{| c | c | c | c | c | c | c | c | c | c |}
        \hline
        \ $n$ \ & \ 2 \ & \ 3 \ & \ 4 \ & \ 5 \ & 6 & 7 & 8 & 9 \\
        \hline
        \ $N_n$ \  & 1 & 2 & 6 & 21 & 112 & 853 & 11,117 & 261,080\\
        \hline
    \end{tabular}
    \caption{The number of connected non-isomorphic graphs $N_n$ grows with the number of vertices $n$.}
    \label{graph count table}
\end{table}

\subsection{Global optimal solutions for $p=1$} \label{couenne section}

We first optimize instances of QAOA for depth $p=1$ using the expectation value of the cost function operator
%
\begin{equation} \label{cost expectation maxcut}\langle \hat C(\gamma_1, \beta_1)\rangle = \sum_{\langle j,k\rangle} \langle \hat C_{\langle j,k\rangle} (\gamma_1, \beta_1)\rangle\end{equation}

\noindent Previously, Hadfield derived a general expression for $\langle \hat C_{\langle j,k\rangle} \rangle$ at depth $p=1$ as \cite{Hadfield2018dissertation} 
$$
\langle \hat C_{\langle j,k\rangle}(\gamma_1, \beta_1) \rangle=  \frac{\sin(4 \beta_1) \sin(\gamma_1) \left(\cos^d(\gamma_1) + \cos^e(\gamma_1)\right)}{4} $$ \begin{equation} \label{Hadfield} -\frac{\sin^2(2\beta_1) \cos^{d+e-2f}(\gamma_1) \left(1-\cos^f(2\gamma_1)\right)}{4} + \frac{1}{2},
\end{equation}

\noindent where $d$ = deg$(j)-1$ and $e$ = deg$(k)-1$ are one less than the degrees of the vertices $j$ and $k$ connected by the edge $\langle j,k\rangle$ and $f$ is the number of triangles containing $\langle j, k\rangle$.

We maximize $\langle \hat C(\gamma_1, \beta_1)\rangle$ using the numerical optimization software Couenne \cite{belotti2009couenne, belotti2009branching}, which optimizes problems of the form 
\begin{align}\label{eq:co} \notag
& \max \  F(\mathbf x, \mathbf y) \\ \notag
& \ \mbox{s.t. }  p_i (\mathbf x,\mathbf y) \leq 0 & \forall i \in [n] \\ \notag
& \ \mathbf x \in \mathbb{R}^n \\
& \ \mathbf y \in \mathbb{Z}^n
\end{align}

\noindent Here, we specify the continuous variable $\mathbf x = (\gamma_1,  \beta_1)$ and the integer variable $\mathbf y$ is not used. The function $F(\mathbf x,\mathbf y)$ then corresponds with Eq.~(\ref{cost expectation maxcut}) while
the polynomial constraints $p_i(\mathbf x)$ are not included since there are no constraints on the angles. Couenne simplifies this problem by reformulating it with auxiliary variables. Then, a convex relaxation of the reformulated problem is found and solved using branch and bound techniques. The process is repeated until an optimal solution is recovered \cite{belotti2009couenne, belotti2009branching}. The results provide globally optimal instances of QAOA for $p=1$.   

\subsection{Numerical searches for general $p$} \label{BFGS section}

\begin{figure*}
    \includegraphics[width=200cm,height=5cm,keepaspectratio,angle=0,trim={0cm 0 1.5cm 0},clip]{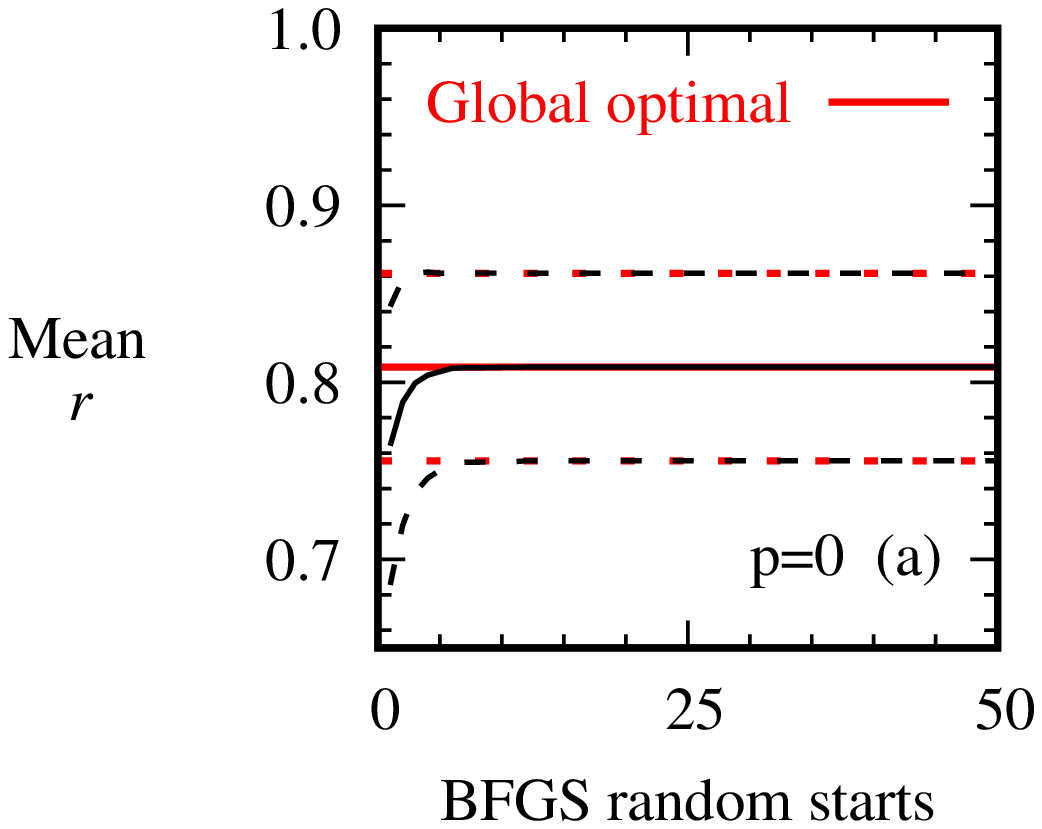}
     \includegraphics[width=200cm,height=5cm,keepaspectratio,angle=0,trim={2.75cm 0 2cm 0},clip]{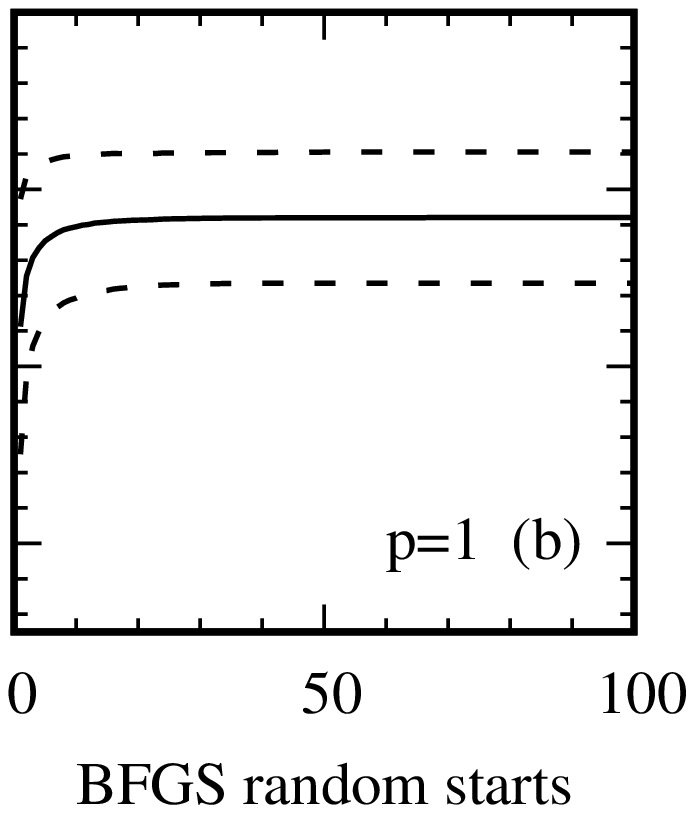}
      \includegraphics[width=200cm,height=5cm,keepaspectratio,angle=0,trim={2.75cm 0 2cm 0},clip]{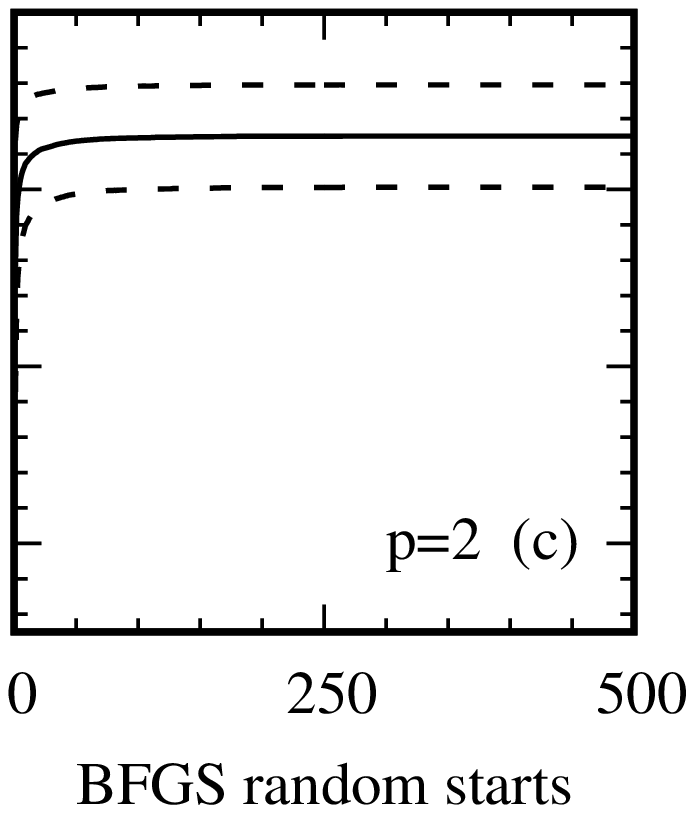}
    \caption{Convergence of multi-start BFGS optimization for the 853 graphs with $n=7$ vertices and depths $p=1,2,3$ in panels (a),(b),(c) respectively. Solid black lines show the mean approximation ratio $r$ over the graphs and black dashed lines show $\pm$ one standard deviation.  For $p=1$ BFGS converges to the global optimal solutions from Couenne in red.}
    \label{BFGS convergence fig}
\end{figure*}

For QAOA depth $p > 1$, we use the Broyden-Fletcher-Goldfarb-Shanno (BFGS) optimization algorithm \cite{NumericalRecipesBFGS}. BFGS has been used previously in a wide variety of contexts including QAOA \cite{zhou2020quantum}.  BFGS begins with an initial set of angles and then iteratively finds angles that converge to a local maximum of $\langle \hat C  (\bm \gamma, \bm \beta) \rangle$.  Each iteration is determined by the numerical gradient of $\langle \hat C \rangle$ and an approximate Hessian second-derivative matrix that is constructed from the gradient at successive steps. Using the approximate Hessian when calculating the steps gives faster convergence than first-order methods while also avoiding the computational expense of calculating the exact Hessian.
 
We test convergence of BFGS using random initial angles and monitoring the local maximum of $\langle \hat C \rangle$.  We repeat the initialization procedure a fixed number of times that varies with depth $p$. We use 50 random-angle seeds when $p=1$, 100 random seeds when $p=2$, and 500 seeds when $p=3$. As shown in Fig.~\ref{BFGS convergence fig}(a), the average and standard deviation of the quantum approximation ratio at $p=1$ quickly converges to the solution obtained using Couenne.
We compared results from BFGS to results from Couenne for every graph with $n \leq 8$ vertices and found exact agreement out to the 6-digit precision of Couenne. We were unable to verify solutions at $n=9$ due to numerical issues with Couenne, so instead we checked these results using brute force searches.  As shown in Fig.~\ref{BFGS convergence fig}(b) and (c), we observe similar behavior in convergence for the cases of $p=2$ and $3$.

We extend our validation of BFGS for $n=9$ with $p=1$ and $n \leq 6$ at $p=2$ using brute force search. We evaluate $\langle \hat C \rangle$ for all  $(\gamma_1,\gamma_2,\beta_1,\beta_2)$ on a grid with spacing $\pi/100$ for each angle. We found both BFGS and brute force return values for the optimal $\langle \hat C \rangle$ to within an additive factor of $10^{-2}$. As BFGS consistently recovers larger maxima, we attribute these differences to coarse-graining in the brute force method. We conclude that BFGS is finding globally optimal results at $p=2$ for these graphs.  

To verify our results are consistent for $n > 6$ with $p=2$ and for all $n$ with $p=3$, we run BFGS calculations a second time using different sets of 100 or 500 random seeds.  We observe an increases in $\langle \hat C \rangle$ after the second round of BFGS for less than $1\%$ of graphs at each $n$ and $p$.  We conclude that BFGS typically finds  globally optimal solutions at these higher $n$ and $p$ though it is unclear how deviations may grow beyond $p=3$.

\subsection{Symmetry analysis}\label{angle symmetry section}

The optimized angles recovered by BFGS exhibit a variety of symmetries, in which multiple angle solutions give the same optimized expectation value of the cost function.  We simplify these results by systematically reporting a single assignment for the optimized angles to compare the distributions of angles recovered across different graph instances. Symmetry analysis proves essential for revealing patterns in the optimized angles discussed in Secs.~\ref{angle section}-\ref{median angle section}.  We summarize how we use the symmetries below with detailed descriptions of the symmetries deferred to the Appendix.

Many of these observed symmetries are identified by extending the analysis of Zhou \etal for regular graphs \cite{zhou2020quantum}, which we apply more broadly to graphs where each vertex has even degree or each vertex has odd degree. Specifically, the angles $\beta_q$ are periodic over intervals of $\pi/2$ and, without loss of generality, may always transform to the interval $-\pi/4 \leq \beta_q \leq \pi/4$ for all $q$.  We always set $\beta_1 < 0$ following the ``time-reversal" symmetry that follows from the dynamics in Eq.~(\ref{psi p}).  There are no symmetries in $\gamma_q$ for generic graphs, so generally these have $-\pi \leq \gamma_q \leq \pi$ for all $q$.  For graphs where every vertex has even degree or every vertex has odd degree, we use symmetries to always transform to angles in the half-intervals $-\pi/2 \leq \gamma_q \leq \pi/2$.   A small number of graphs have additional symmetries and for these we report angles with the most component $(\gamma_q,\beta_q)$ pairs inside the intervals
\begin{equation} \label{GammaBeta} \Gamma = [-\pi/2,0], \ \ \Beta=[-\pi/4,0]. \end{equation}

\section{QAOA Simulation Results} \label{results}

We analyze QAOA performance on the MaxCut problem for graphs with various numbers of vertices $n$ at various depths $p$  using the two performance measures of Section \ref{success measures}.  We also discuss patterns in the observed optimized angle distributions and we construct a search heuristic that uses these patterns to reduce the computational expense of parameter optimization. 

\subsection{Approximation ratio} \label{expectation c section}

\begin{figure}
\includegraphics[width=200cm,height=8.6cm,keepaspectratio,angle=270,trim={0cm 2.cm 1.75cm 4cm},clip]{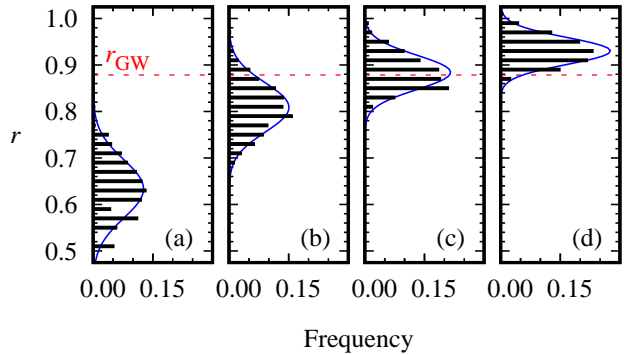}
    \caption{Distributions of the approximation ratios $r$ for graphs with $n=7$ vertices.  Panels (a),...,(d) show $p=0,...,3$ respectively, the red dashed line shows the approximation ratio from the Goemann-Williamson algorithm $r_\mathrm{GW}$.} 
\label{distribution expectation c}
\end{figure}

Our first measure of QAOA performance on MaxCut is the approximation ratio $r$ from Eq.~(\ref{avg r}).  We present the distribution of $r$ across graphs with various $n$ and $p$. Figure \ref{distribution expectation c} shows an example of the distribution of $r$ for all 853 graphs with $n=7$ vertices for depths $p = 0,\ldots,3$ in panels (a)-(d), respectively. Note that $p=0$ corresponds to the initial state in Eq.~(\ref{initial state}).  Trends for other $n$ are similar and described further  below. In Fig.~\ref{distribution expectation c}, the width of the plotted bars is  smaller than the histogram bin width $w=0.02$ to clearly visualize the distributions.  Each panel also overlays a Gaussian distribution $(w/\sqrt{2\pi\sigma^2})\exp(-(r-\bar r)^2/2\sigma^2)$ using the mean $\bar r$ and standard deviation $\sigma$ in Table \ref{expectation c table} as calculated from the observed distributions of $r$.  
 
 \begin{table}
    \centering
    \begin{tabular}{| c | c | c | c | c | c |}
        \hline
        $\ p\ $ &\ $\bar r$ &\ $\sigma$\ &\  $\bar F_{0.7} $\ & \  $\bar F_{0.8} $\ & \ $\bar  F_{0.9} $\ \\     
         \hline
        0 &    $ 0.633$ & $0.063 $   &  0.116 & 0.004 & 0.000 \\
        \hline
        1 & $ 0.809$ & $ 0.053 $   & 0.986 & 0.551 & 0.039\\
        \hline
        2 & $ 0.884$ & $ 0.037 $   & 1.000 & 1.000 & 0.321  \\
        \hline
        3 & $  0.930$ & $ 0.029 $  & 1.000 & 1.000 & 0.825 \\
        \hline
    \end{tabular}
    \caption{Mean $\bar r$, standard deviation $\sigma$, and complimentary cumulative distribution functions $\bar F_R(r)$ of Eq.~(\ref{cumulative dist function}) for the $r$ distributions in Fig.~\ref{distribution expectation c}.}
    \label{expectation c table}
\end{table}

 \begin{figure*}
    \includegraphics[width=20cm,height=5.25cm,keepaspectratio,angle=0,trim={1cm 0 0cm 0},clip]{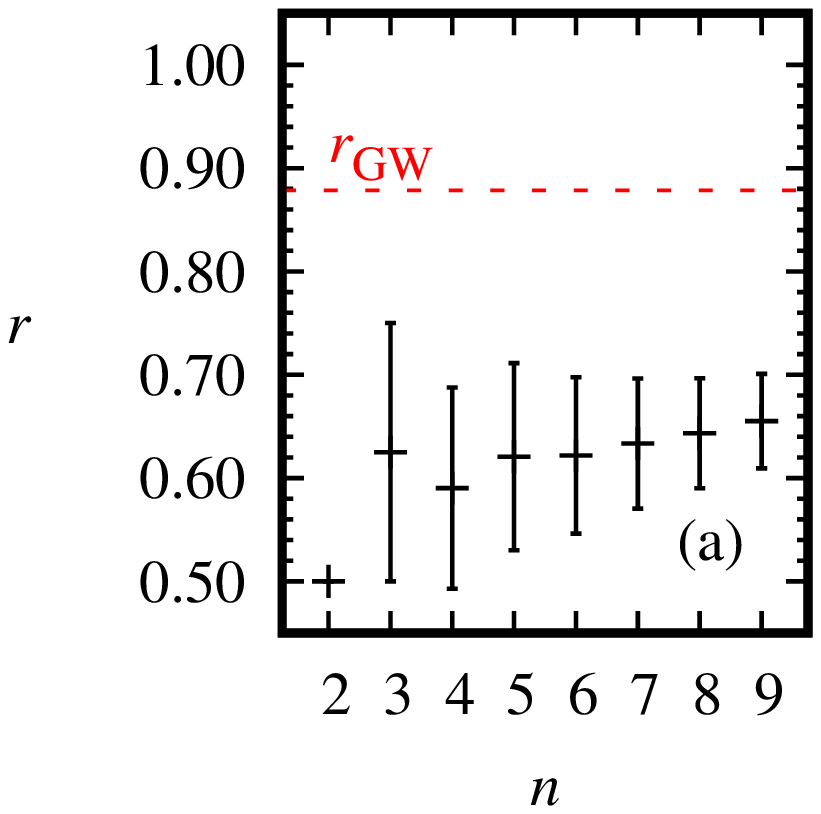}
    \includegraphics[width=20cm,height=5.25cm,keepaspectratio,angle=0,trim={4cm 0cm 0cm 0},clip]{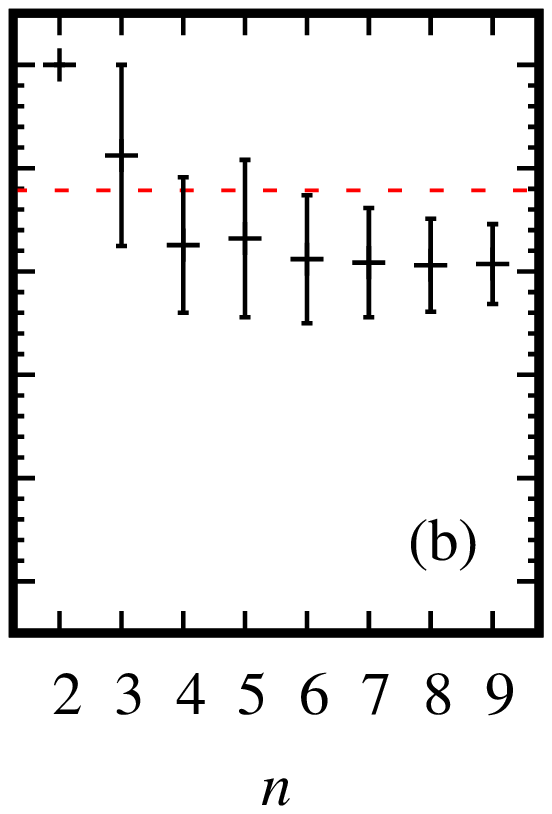}
     \includegraphics[width=200cm,height=5.25cm,keepaspectratio,angle=0,trim={4cm 0cm 0cm 0},clip]{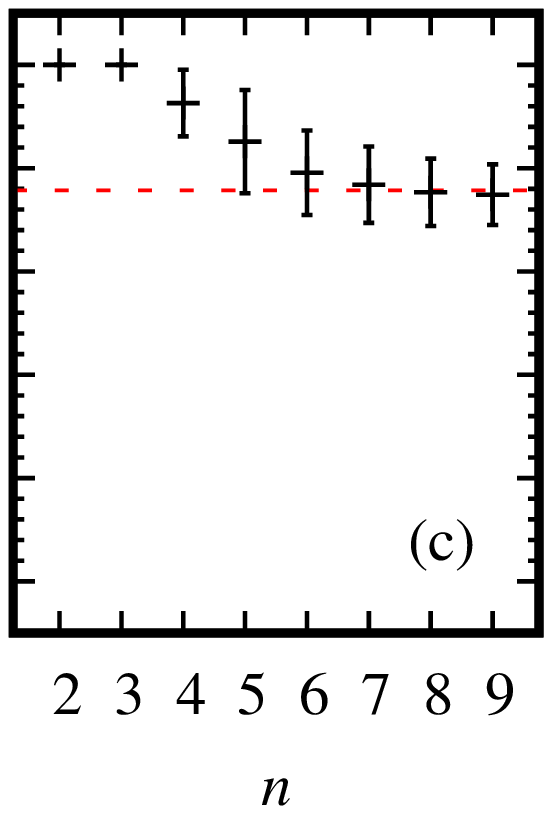}
      \includegraphics[width=200cm,height=5.25cm,keepaspectratio,angle=0,trim={4cm 0cm 0cm 0},clip]{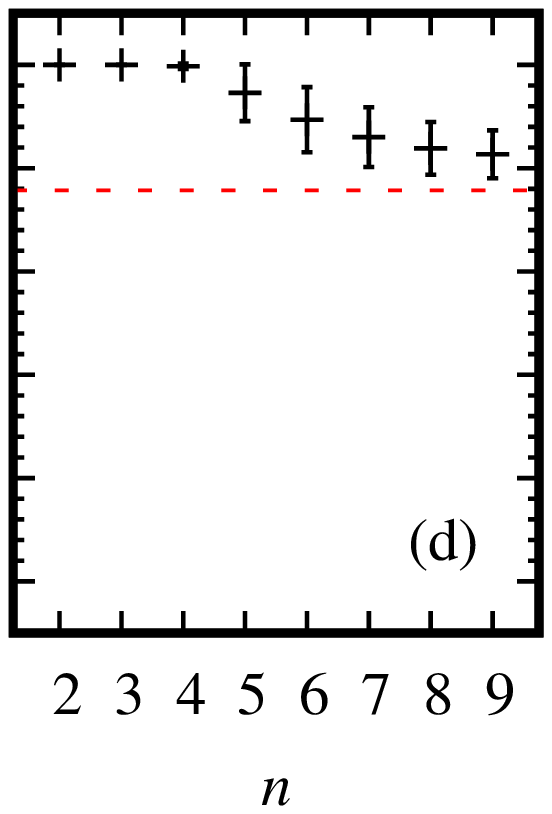}
    \caption{Mean $r$ across all graphs with various numbers of vertices $n$ and depths $p$.  Panels (a),...,(d) show $p=0,...,3$ respectively.  Error bars represent standard deviations of the $r$, the dashed red line represents the approximation ratio $r_\mathrm{GW}$ of the Goemans-Williamson algorithm.  }
    \label{expectation c fig}
\end{figure*}

Figure \ref{distribution expectation c}(a) shows $r$ for the initial states of Eq.~(\ref{initial state}).  Every $z$ is equally probable in these states, so $r$ is equivalent to the approximation ratio obtained by averaging random cuts from the uniform distribution.  This cuts half the edges in $E$ on average, $\langle \hat C \rangle = E/2$, while the maximum number of cuts is bounded by the total number of edges, $C_\mathrm{max} \leq E$, so $r = \langle \hat C\rangle/C_\mathrm{max} \geq 1/2$. 

Figures \ref{distribution expectation c}(b), (c), and (d)  show similar distributions of the $r$ for $p=1,2,3$ respectively.  These are well described by Gaussian distributions for each $p$ and $r$ is approaching one as $p$ increases. QAOA instances also perform more similarly with increasing depth as indicated by the decrease in standard deviation of  $r$.

We further quantify increases in $r$ with $p$ by calculating $\bar F_R(r)$ as the fraction of graphs with approximation ratio $r$ that exceeds a threshold $R$.  This is given by the complimentary cumulative distribution function 
 \begin{equation} \label{cumulative dist function}  \bar F_R(r ) = \frac{1}{N_{n}}\sum_{G(n)} \Theta(r - R),  \end{equation} 

\noindent where $N_n$ is the number of non-isomorphic connected graphs with $n$ vertices, $\sum_{G(n)}$ is the sum over these graphs, and $\Theta$ is the Heaviside step function. Table \ref{expectation c table} presents $\bar F_R(r )$ for $n=7$ with $R = 0.7, 0.8,$ and $0.9$. Almost all graphs exceed $r \geq 0.7$ at $p=1$ and more than half exceed $r \geq 0.8$.  At $p=2$, all graphs have $r > 0.8$ while most graphs have $r > 0.9$ at $p=3$. We note that the latter results exceeds the best lower bound for the approximation ratio of $r_{GW} = 0.87856$ set by the Goemans-Williamson algorithm \cite{GWalgorithm}.
 
We analyzed the distribution of $r$ for all $n\leq 9$.  Similar trends are observed for all $n$: the mean $r$ increases as $p$ increases and the distributions become narrower in terms of the standard deviation. The distributions of $r$ over different graphs at the same $n$ are well described by Gaussian distributions when $n \geq 7$, while for $n < 7$ the small numbers of graphs yield more irregular histogram distributions.
 
Figure~\ref{expectation c fig} shows the mean of $r$ as a function of $n$ for $p=0,\ldots,3$ in panels (a)-(d), respectively. Averages for $p=0$ are approximately 2/3 for all $n$ while the standard deviation decreases with increasing $n$ due to the number of graphs sampled increasing as $N_n$. For $p\neq0$ in Figs.~\ref{expectation c fig}(b)-(d), the smallest graphs reach an optimal $r=1$ while larger graphs decay to steady values of $r<1$ as $n$ increases.  There is not much variation in the mean or standard deviation of $r$ between nearby $n$, especially at the largest $n$, so larger $n$ might be expected to return similar $r$ distributions.  As $p$ increases, at each $n$ the mean $r$ become larger and the standard deviations become smaller, which is consistent with the trends at $n=7$ seen in Fig.~\ref{distribution expectation c}. The mean $r$ all exceed the Goemans-Williamson $r_\mathrm{GW}$ by at least a standard deviation at $p=3$.

\subsection{Probability of the maximum cut} \label{pcmax section}

\begin{figure}
    \includegraphics[width=200cm,height=8.6cm,keepaspectratio,angle=270,trim={0.7cm 2.25cm 1.75cm 4.4cm},clip]{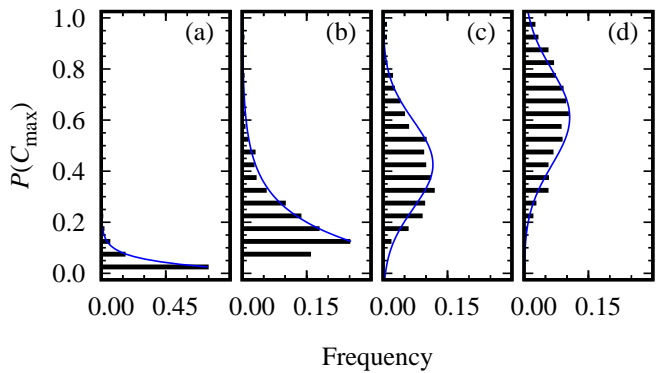}
    \caption{Distributions of probabilities  to obtain the maximum cut $P(C_\mathrm{max})$ for graphs with $n=7$ vertices.  Panels (a),..,(d) show $p=0,...,3$ respectively.}
    \label{distribution pcmax fig}
\end{figure}

We next assess the performance of QAOA in terms of the probability for obtaining the maximum cut, $P(C_\mathrm{max})$. Figure \ref{distribution pcmax fig} shows an example of the normalized distribution of the $P(C_\mathrm{max})$ for the 853 graphs with $n=7$ at varying depths $p$. Statistics describing these distributions are given in Table \ref{pcmax table}.  

\begin{table}
    \centering
    \begin{tabular}{| c | c | c | c | c | c |}
        \hline
         $\ p\ $ &\ $\bar P(C_\mathrm{max})$\ &\ $\sigma$\ &\  $\bar F_{0.25} $\ & \  $\bar F_{0.50} $\ & \ $\bar F_{0.75} $\ \\       
          \hline
        0 &    $ 0.046$ & $0.051$ & 0.009 & 0.001 & 0.000 \\
        \hline
        1 & $ 0.211 $ & $ 0.132 $ & 0.278 & 0.035 & 0.009  \\
        \hline
        2 & $ 0.426$ & $ 0.172 $ & 0.835 & 0.318 &  0.046  \\
        \hline
        3 & $0.610$  & $ 0.187$ &  0.975 & 0.715 &  0.252 \\
        \hline
    \end{tabular}
    \caption{Mean $\bar P(C_\mathrm{max})$, standard deviation $\sigma$, and complimentary cumulative distribution functions $\bar F_R(P(C_\mathrm{max}))$ similar to Eq.~(\ref{cumulative dist function}) for the $P(C_\mathrm{max})$ distributions in Fig.~\ref{distribution pcmax fig}.} 
    \label{pcmax table}
\end{table}

The distribution for $p=0$ shown in  Fig.~\ref{distribution pcmax fig}(a) peaks around $P(C_\mathrm{max}) \approx 0$ with a rapid decrease away from zero, as expected. An exponential fit for this histogram distribution yields $\exp(-\kappa_0P(C_\mathrm{max}))$ with $\kappa_0  = 29.1 \pm 0.9$. At $p=0$, each possible cut is equally probable in the initial state and there are $2^n$ possible cuts with at least two maximum cuts. This implies the bound
\begin{equation} \label{p min} P_0(C_\mathrm{max}) \geq 1/2^{n-1},\end{equation} 

\noindent where the ``0" subscript refers to the initial state.  The minimum value $P_0(C_\mathrm{max})=1/2^{n-1}$ gives the main contribution to the lowest bar in Fig.~\ref{distribution pcmax fig}(a) while higher values are obtained for those graphs with additional symmetries, for example, in a complete graph where symmetry gives $n!/(\lfloor n/2 \rfloor!\lceil n/2 \rceil!)$ maximum cuts.  The peak around $P_0(C_\mathrm{max})=1/2^{n-1}$ indicates that large degeneracies like this are uncommon in the total set of graphs.

Figure \ref{distribution pcmax fig}(b) shows $P(C_\mathrm{max})$ when $p = 1$.  We fit this distribution with an exponential starting at the second histogram bar and similar to the $p=0$ distribution. This yields a smaller decay constant $\kappa_1 = 7.0 \pm 0.3$.  Panels (c) and (d) show the distributions for $p=2$ and $3$, respectively.  These distributions are better described by Gaussian distributions with means and standard deviations presented in Table \ref{pcmax table}. The means increase with $p$ and the distributions widen indicating greater differences between graphs at larger $p$. The graphs pass thresholds $P(C_\mathrm{max}) > R$ gradually, as seen by  $\bar F_R(P(C_\mathrm{max}))$.

We contrast distributions of $P(C_\mathrm{max})$ against distributions of the quantum approximation ratio $r$ in Fig.~\ref{distribution expectation c} and Table \ref{expectation c table}.  The mean of $P(C_\mathrm{max})$ increases the least when going from $p=0$ to $p=1$, while by contrast the mean of $r$ increases the most in going from $p=0$ to $p=1$. The exponential and Gaussian distributions are also significantly different at small $p$. At higher $p$, $P(C_\mathrm{max})$ and $r$ both follow Gaussian distributions but there are differences in how their widths change with $p$, with increasing differences between different graphs for $P(C_\mathrm{max})$ and the opposite for $r$.  

\begin{figure}
   \input{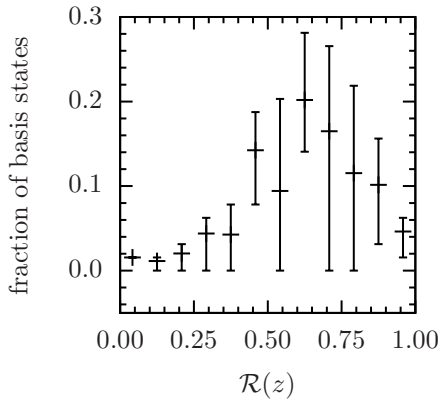}
    \caption{Average spectrum of the normalized cost function $\mathcal{R}(z)$ from Eq.~(\ref{rz}) for graphs with $n=7$ vertices.}
    \label{cz spectrum}
\end{figure}

We explain why distributions of $r$ have high averages and narrow widths when concurrently $P(C_\mathrm{max})$ presents broad widths and relatively small averages.  Consider the normalized eigenspectrum $\hat C/C_\mathrm{max}$ of energies $\mathcal{R}(z)=C(z)/C_\mathrm{max}$.  For each $n$, we made a histogram of $\mathcal{R}(z)$ for each graph, then average these values to obtain an average spectrum for $\mathcal{R}(z)$. Figure \ref{cz spectrum} shows the average spectrum of $\mathcal{R}(z)$ at $n=7$ vertices with each data point indicating the fraction of basis states in a histogram bin of width 1/12 averaged over all graphs with $n=7$.  The results look similar for other $n \geq 6$ while  the distributions for $n \leq 5$ are irregular due to small numbers of graphs. We show error bars in the figure denoting the first and third quartiles of the distributions of the fractions of basis states, calculated by listing the fractions for each graph in increasing order then taking the entries 1/4 and 3/4 of the way up the list, rounded to the nearest integer. 

Most basis states give sub-optimal cuts $\mathcal{R}(z) < 1$ in the average spectrum of Fig.~\ref{cz spectrum}, with the majority of states in the range $2/3 \lesssim \mathcal{R}(z) \lesssim 0.95$ and with few optimal states at $\mathcal{R}(z) = 1$. 
Thus, optimization of $r$ in state preparation favors large total probabilities in many states with near-optimal $\mathcal{R}(z)$ over the smaller probability of preparing truly optimal states.  This gives relatively high and uniform $r$ for different graphs but a much more variable $P(C_\mathrm{max})$.

\begin{figure*}
      \includegraphics[width=20cm,height=5.25cm,keepaspectratio,angle=0,trim={0cm 0 0cm 0},clip]{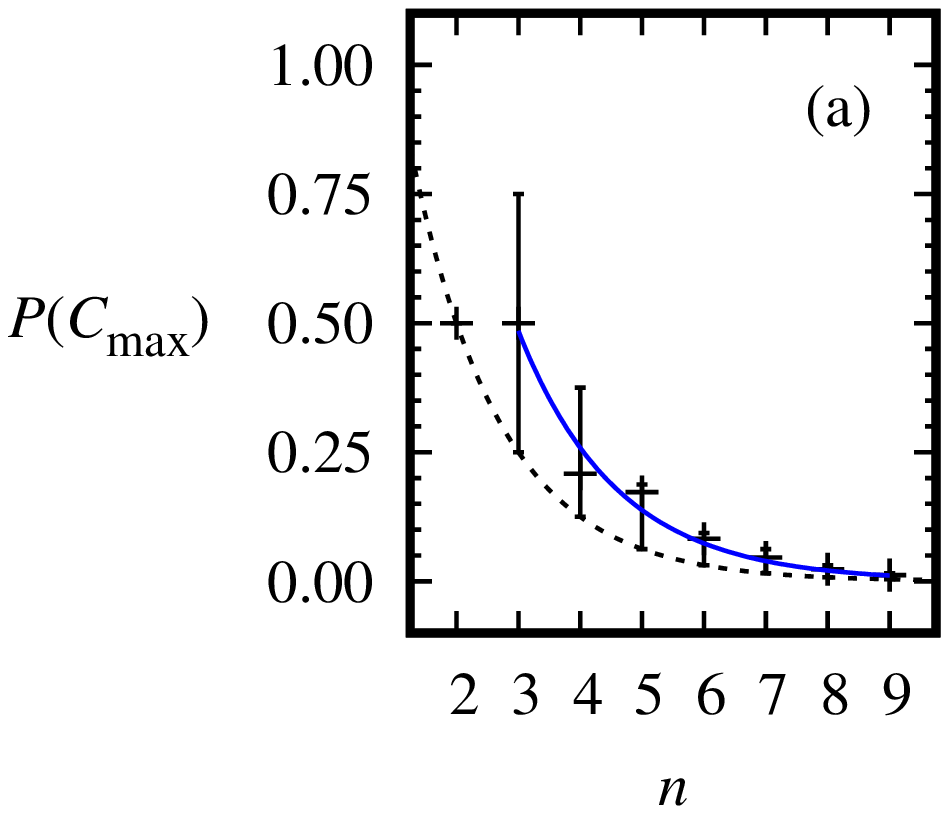}
    \includegraphics[width=20cm,height=5.25cm,keepaspectratio,angle=0,trim={4cm 0cm 0cm 0},clip]{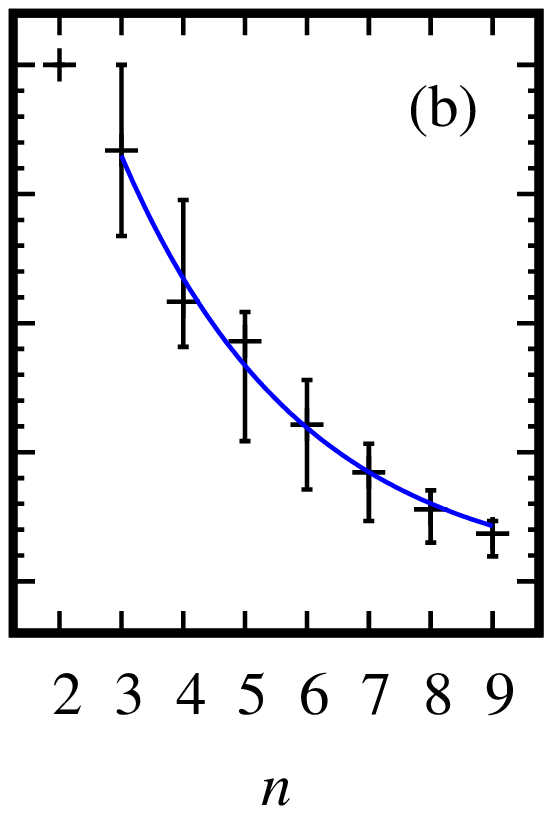}
     \includegraphics[width=20cm,height=5.25cm,keepaspectratio,angle=0,trim={4cm 0cm 0cm 0},clip]{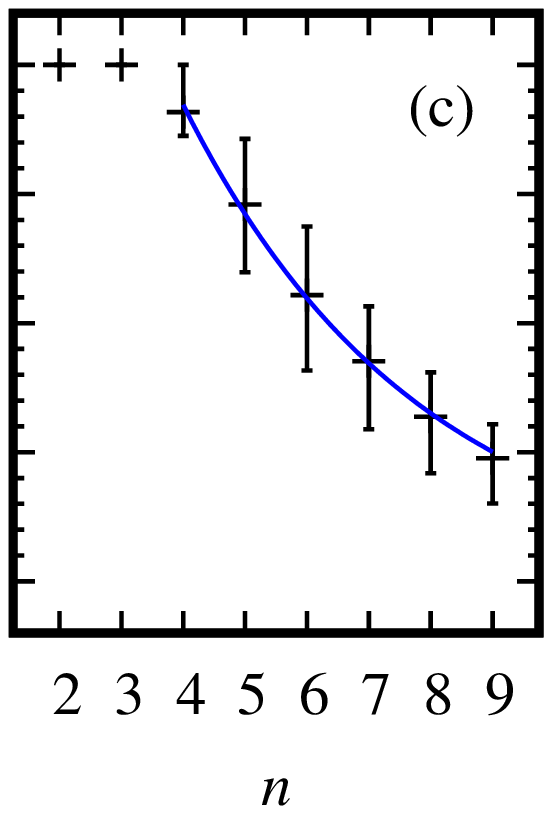}
     \includegraphics[width=20cm,height=5.25cm,keepaspectratio,angle=0,trim={4cm 0cm 0cm 0},clip]{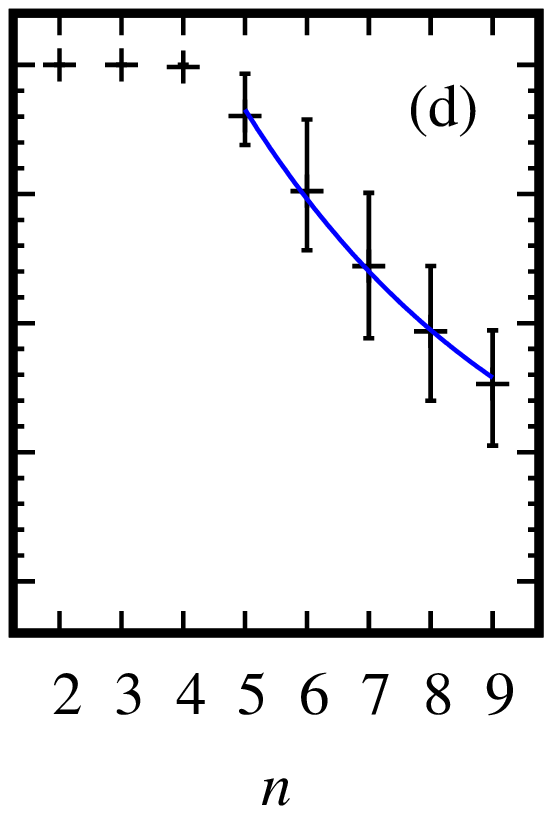}
    \caption{Probabilities for obtaining the maximum cut $P(C_\mathrm{max})$, averaged over graphs at various $n$, with $p=0,...,3$ in panels (a),...,(d) respectively.  Asymmetric error bars show the quartiles of the distributions of $P(C_\mathrm{max})$, the dashed black line in (a) follows $1/2^{n-1}$ from Eq.~(\ref{p min}), solid blue lines show the exponential fits of Eq.~(\ref{Pfit}) to the indicated data, with parameters from Table \ref{pcmax exp fit table}.}
    \label{expectation pcmax fig}
\end{figure*}

Figure \ref{expectation pcmax fig} shows how the average of $P(C_\mathrm{max})$ varies for the distribution of graphs at various $n$ and $p$ using asymmetric error bars expressed as quartiles. The mean of $P(C_\mathrm{max})$ decreases exponentially with $n$ as shown by the blue fit to the averages 
\begin{equation} \label{Pfit} P_\mathrm{mean}^{\mathrm{fit}}(C_\mathrm{max}) = N_pe^{-k_pn} \end{equation}

\noindent with the fit parameters in Table \ref{pcmax exp fit table}.  We fit to a subset of the data at each $p$ to include only the largest $n$ where decay is observed, and we show the $n$ we fit at each $p$ by the location of the fit curve. 

At $p=0$ the first quartiles follow the minimum of Eq.~(\ref{p min}), shown by the dashed black line.  This indicates many of the graphs saturate the minimum theoretical value of $P_0(C_\mathrm{max})$.  As $n$ increases, the averages and quartiles of the $P(C_\mathrm{max})$ distributions approach small values near the minimum of Eq.~(\ref{p min}).

\begin{table}
    \centering
    \begin{tabular}{| c | c | c |}
        \hline
         $\ p\ $ & $N_p$ & $k_p$ \\       
          \hline
         0 & $3.2 \pm 0.8$ & $  0.63 \pm 0.07$ \\
        \hline
        1 & $2.3 \pm 0.2$ & $  0.34 \pm 0.02$ \\
        \hline
        2 & $2.6 \pm 0.1$ &  $0.260 \pm 0.008$  \\
        \hline
        3 & $2.6 \pm 0.2$ & $ 0.210 \pm 0.008$ \\
        \hline
         \end{tabular}
    \caption{Parameters for the exponential fits Eq.~(\ref{Pfit}) in Fig.~\ref{expectation pcmax fig}.}
    \label{pcmax exp fit table}
\end{table}

Figure \ref{expectation pcmax fig}(b) shows the averages and quartiles of the distributions of $P(C_\mathrm{max})$ at $p =1$.  The average $P(C_\mathrm{max})$ have increased relative to $p=0$ and are again decreasing exponentially with $n$, but the rate of exponential decay $k_p$ is about half of the rate at $p=0$, see Table \ref{pcmax exp fit table}. The error bars denoting quartiles reduce for $n=3$, as these graphs are approaching full optimization.  For higher $n$ the error bars increase to indicate there are larger deviations between different graphs at $p=1$.  

The same trends continue at higher $p$: the average $P(C_\mathrm{max})$ increase with $p$, with decreasing exponential decay constants in Table \ref{pcmax exp fit table}.  The distributions widen as $p$ increases at large $n$, indicating $P(C_\mathrm{max})$ is increasingly sensitive to graph structure.  This is in contrast to the distributions of the $r$, which narrow as $p$ increases, with more similar $r$ for different graphs.  

\subsection{Optimized angle distributions}  \label{angle section}

We next discuss the distribution of optimized angle parameters shown in Figs.~\ref{p=1 angle fig}-\ref{p=3 angle fig} for the 853 graphs at $n=7$.  These figures present two-dimensional histograms of the angle distributions, where the optimized ($\gamma_j, \beta_j)$ have been organized into bins of size $\pi/20\times\pi/20$ and counted.  We use a logarithmic color scale to visualize the distributions and add a base of one counts in each bin so the logarithm does not diverge when there are zero counts.  

Patterns have been observed previously in the optimized angles for solving simple families of graphs.  Optimized angle patterns have been observed for 3-regular graphs \cite{zhou2020quantum}, small samples of random graphs \cite{crooks2018performance,Shaydulin2020CaseStudy}, and even in a variant formulation of QAOA solving the Max-$k$ Vertex Cover problem \cite{Bartschi2020MaxkCover}.  An argument has been given that angle patterns should be expected within families of graphs with systematic structure \cite{brandao2018concentration}.  However, it is not clear if similar angle patterns hold for over the set of general graphs considered here.

\subsubsection{Angle patterns at $p=1$}

\begin{figure}
    \centering
    \includegraphics[width=9cm,height=8cm,keepaspectratio,angle=0,trim={0 2.5cm 0 2cm},clip]{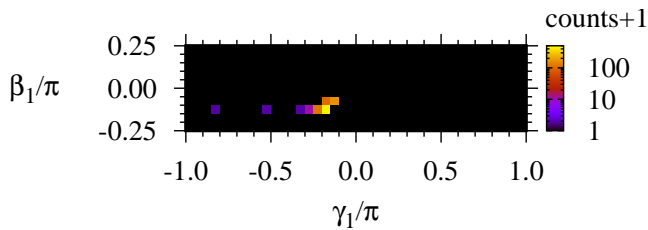}
    \caption{Optimized angle distribution histogram for $p=1$ QAOA on graphs with $n=7$ vertices.}
    \label{p=1 angle fig}
\end{figure}

Figure \ref{p=1 angle fig} shows the distribution of optimized angles for all graphs with $n=7$ vertices at depth $p=1$ in Eq.~(\ref{psi p}).  The overwhelming majority of the angles are focused in the bright spot near $\gamma_1 \approx -\pi/6$, $\beta_1 \approx -\pi/8$.  This concentration is generic for all $n$ studied: a very limited range of angles $(\gamma_1, \beta_1)$ are observed to optimize almost all graphs with $p=1$.  

We categorize the optimized angles using a partitioning of parameter space. Using $\Gamma = [-\pi/2,0]$ and  $\Beta = [-\pi/4,0]$ from Eq.~(\ref{GammaBeta}), we say angle-pairs $(\gamma_j,\beta_j)$ inside $(\Gamma,\Beta)$ follow the angle patterns and say angle-pairs outside $(\Gamma,\Beta)$ deviate from the pattern.  Categorizing by $(\Gamma,\Beta)$ quantifies how many graphs have angles that follow the observed patterns in Fig.~\ref{p=1 angle fig} at $p=1$ and also extends to categorizing graphs when $p>1$.

Let $D_{n,p}$ be the number of $n$-vertex graphs that deviate from the angle patterns at depth $p$.  At $p=1$, $D_{n,p}$ is the number of graphs with $(\gamma_1,\beta_1) \not\in(\Gamma,\Beta)$. For general $p$, $D_{n,p}$ is the number of graphs with $(\gamma_j,\beta_j) \not\in(\Gamma,\Beta)$ for any $j$.  Let 
\begin{equation} \label{deviations} \mathcal{D}_{n,p} = \frac{D_{n,p}}{N_n} \end{equation}

\noindent be the fraction of $n$-vertex graphs that deviate from the patterns at depth $p$, where the $N_n$ is the total number of graphs from Table \ref{graph count table}. We calculate the $D_{n,p}$ using a numerical error tolerance of $\epsilon = 10^{-5}$, so that $\gamma_j$ that are in $\Gamma$ to within numerical error are counted as $\gamma_j \in \Gamma$, and similarly for the $\beta_j$.

Table \ref{p=1 angle table} lists $D_{n,p}$ and the fractions $\mathcal{D}_{n,p}$ of graphs that do not follow the angle patterns at $p=1$ for $n \leq 9$.  Every graph follows the pattern with $n \leq 6$, a small number of graphs deviate beginning with $n=7$.  The number of graphs that deviate  increases with $n$, but they are a small fraction of the total number of graphs.

\begin{table}
    \centering
    \begin{tabular}{| c | c | c | c |}
        \hline
         \ $n$ \ & \ $p$ \ & \ $D_{n,p}$ \ & \ $\mathcal{D}_{n,p}$ \ \\       

          \hline
        2 &  1 & 0 & 0 \\
        \hline
        3 &  1 & 0 & 0 \\
        \hline
        4 & 1 & 0 & 0 \\
        \hline
        5 & 1 & 0 & 0  \\
        \hline
        6 & 1 & 0 & 0\\
        \hline
        7 & 1 & 1 & $1.172 \times 10^{-3}$\\
        \hline
        8 & 1 & 1 & $8.995 \times 10^{-5}$ \\
        \hline
        9 & 1 & 89 & $3.409 \times 10^{-4}$\\
        \hline
    \end{tabular}
    \caption{Numbers $D_{n,p}$ and fractions $\mathcal{D}_{n,p}$ of graphs with optimized angles $(\gamma_1,\beta_1)\not\in(\Gamma,\Beta)$ at $p=1$.}
    \label{p=1 angle table}
\end{table}

\subsubsection{Angle patterns at $p > 1$}

\begin{figure}
    \includegraphics[width=9cm,height=8cm,keepaspectratio,angle=0,trim={0 2.25cm 0 2cm},clip]{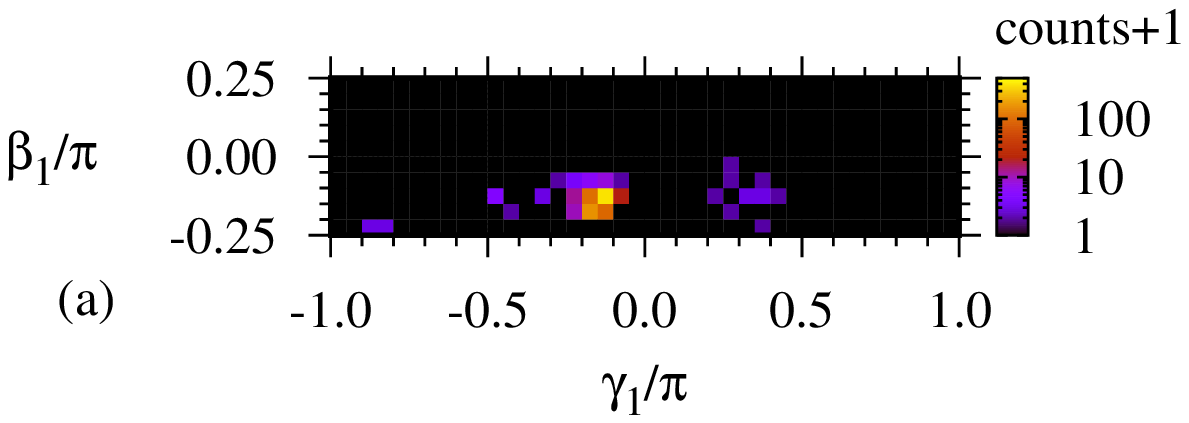}
    \includegraphics[width=9cm,height=8cm,keepaspectratio,angle=0,trim={0 2.5cm 0 2.5cm},clip]{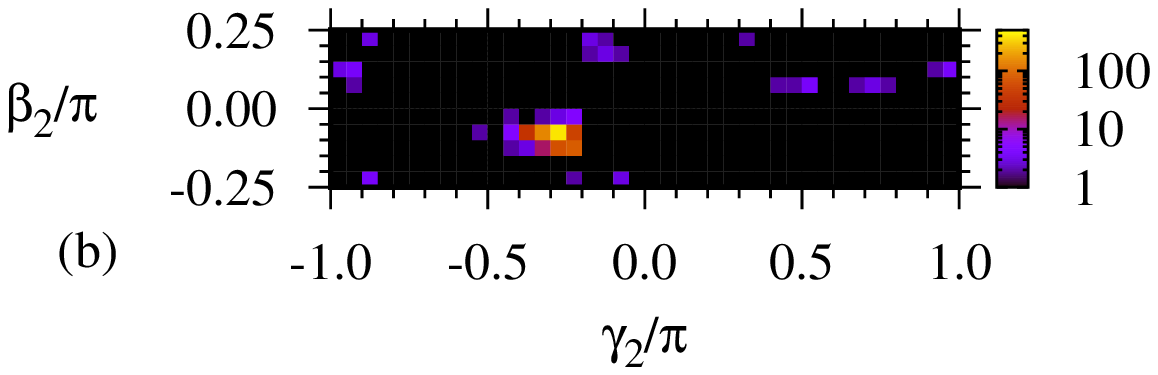}
    \caption{Angle patterns at depth $p=2$ for graphs with $n=7$ vertices, with $(\gamma_1,\beta_1)$ from the first layer in (a) and $(\gamma_2,\beta_2)$ in (b).}
    \label{p=2 angle fig}
\end{figure}

Figure \ref{p=2 angle fig} shows the distributions of optimized angles for all graphs with $n=7$ vertices with depth $p=2$.   The observed $(\gamma_1, \beta_1)$ differ from the values at $p=1$ because these angles are optimized together with $(\gamma_2,\beta_2)$ at $p=2$.  The majority of $(\gamma_1, \beta_1)$ are still concentrated near $\gamma_1 \approx -\pi/6$, $\beta_1 \approx -\pi/8$ as for $p=1$, but the distribution is more dispersed within $(\Gamma,\Beta)$ in comparison with Fig.~\ref{p=1 angle fig}.  There is a new cluster of angles at  $\gamma_1 \approx \pi/3$ with a spread over all the $\beta_1$ (recall $\beta_1 \leq 0$ by the symmetry of Section \ref{angle symmetry section}), and a small cluster of graphs with angles  $\gamma_1 \approx -9\pi/10$, $\beta_1 \approx -\pi/4$.

Figure \ref{p=2 angle fig}(b) shows the distribution of the second layer of angles $(\gamma_2,\beta_2)$.  The majority of angles follow a peaked distribution in the parameter space of $(\Gamma,\Beta)$.  The distribution is shifted from the distribution at $p=1$ to new angles $\gamma_2 \approx -\pi/4$ and $\beta_2 \approx -\pi/13$.  There are also subsets of graphs with angles that form clusters around seemingly-random parameter values.

We again categorize the optimized angles as agreeing with the pattern when they are inside the parameter space of $(\Gamma,\Beta)$ from Eq.~(\ref{GammaBeta}) and deviating from the pattern when they are outside $(\Gamma,\Beta)$ to within numerical error.  The total fraction of graphs that deviate from the pattern is $\mathcal{D}_{n,p}$ from Eq.~(\ref{deviations}).  We further separate the  $\mathcal{D}_{n,p}$ into components to understand which $(\gamma_j,\beta_j)$ deviate and their correlations at $p=2,3$.

Let $\mathcal{D}^{(j)}_{n,p}$ denote the fraction of graphs that deviate only in the $j$th angle pair, $(\gamma_j,\beta_j) \not\in (\Gamma,\Beta)$ and $(\gamma_k,\beta_k) \in (\Gamma,\Beta)$ for all $k \neq j$.  Let $\mathcal{D}^{(jk)}_{n,p}$ denote the fraction of graphs where two angle pairs deviate, $(\gamma_j,\beta_j) \not\in (\Gamma,\Beta)$ and $(\gamma_k,\beta_k) \not\in (\Gamma,\Beta)$ but $(\gamma_l,\beta_l) \in (\Gamma,\Beta)$ for all $l \not\in \{k,j\}$, and similarly let $\mathcal{D}^{(jkl)}_{n,p}$ denote the fraction of graphs where the $j$th, $k$th, and $l$th angle-pairs deviate from $(\Gamma,\Beta)$.    The total fraction of graphs that deviate from the angle patterns is the sum of fractions that deviate in different sets of angle pairs, for example, at $p=2$ the total fraction of graphs that deviate is the sum of fractions that deviate in one or both angle pairs, $\mathcal{D}_{n,p} = \mathcal{D}^{(1)}_{n,p} + \mathcal{D}^{(2)}_{n,p} + \mathcal{D}^{(12)}_{n,p}$.

Table \ref{p=2 angle table} shows the fractions of graphs that deviate from the angle patterns at $p=2$.  The total fraction of graphs that deviate $\mathcal{D}_{n,p}$ typically decreases as $n$ increases, however, the $\mathcal{D}_{n,p}$ is also increasing with $p$ as seen by comparison with Table \ref{p=1 angle table}.  The angles deviate most often in $(\gamma_2,\beta_2)$ or in both $(\gamma_1,\beta_1)$ and $(\gamma_2,\beta_2)$, with $\mathcal{D}^{(2)}_{n,p} \approx \mathcal{D}^{(12)}_{n,p}$, while deviations in only  $(\gamma_1,\beta_1)$ are observed less often, with $\mathcal{D}^{(1)}_{n,p} < \mathcal{D}^{(2)}_{n,p}, \mathcal{D}^{(12)}_{n,p}$.

\begin{table}
    \centering
    \begin{tabular}{| c | c | c | c | c | c |}
        \hline
         \ $n$ \ & \ $p$ \ & $\mathcal{D}^{(1)}_{n,p}$ & $\mathcal{D}^{(2)}_{n,p}$ & $\mathcal{D}^{(12)}_{n,p}$ & $\mathcal{D}_{n,p}$\\       
          \hline
        2 &  2 & 0.000 & 0.000 & 0.000 & 0.000\\
          \hline
        3 &  2 & 0.000 & 5.000$\times 10^{-1}$ & 0.000 & 5.000$\times 10^{-1}$ \\
        \hline
        4 & 2 & 0.000 & 0.000  & 1.667$\times 10^{-1}$ & 1.667$\times 10^{-1}$\\
        \hline
        5 & 2 &  0.000 &  1.429$\times 10^{-1}$ & 4.762$\times 10^{-2}$ & 1.905 $\times 10^{-1}$\\
        \hline
        6 & 2 & 0.000 & 2.679$\times 10^{-2}$ & 8.036$\times 10^{-2}$ & 1.071$\times 10^{-1}$ \\
        \hline
        7 & 2 & 4.689$\times 10^{-3}$ & 2.579$\times 10^{-2}$ & 1.290$\times 10^{-2}$ & 4.338$\times 10^{-2}$ \\
        \hline
        8 & 2 & 4.498$\times 10^{-4}$ &5.847$\times 10^{-3}$ & 4.678$\times 10^{-3}$ & 1.097$\times 10^{-2}$\\
        \hline
        9 & 2 & 4.596$\times 10^{-5}$ & 1.360$\times 10^{-3}$ & 9.652$\times 10^{-4}$ & 2.371$\times 10^{-3}$ \\
        \hline    \end{tabular}
    \caption{Fractions of graphs with optimized angles that deviate from the angle-pattern parameter space $(\Gamma,\Beta)$ at $p=2$, see text for details.}
    \label{p=2 angle table}
\end{table}

\begin{figure}
    \includegraphics[width=9cm,height=8cm,keepaspectratio,angle=0,trim={0 2.25cm 0 2cm},clip]{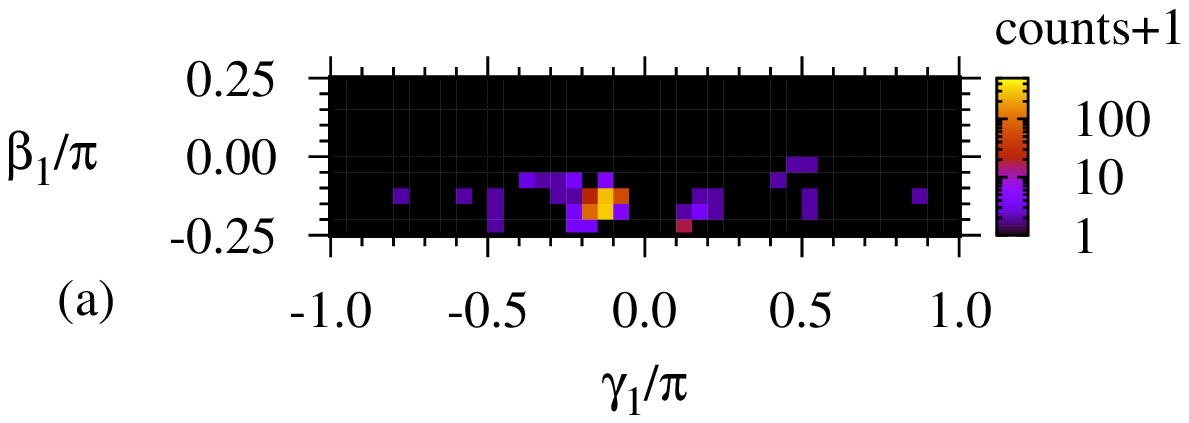}
    \includegraphics[width=9cm,height=8cm,keepaspectratio,angle=0,trim={0 2.25cm 0 2.5cm},clip]{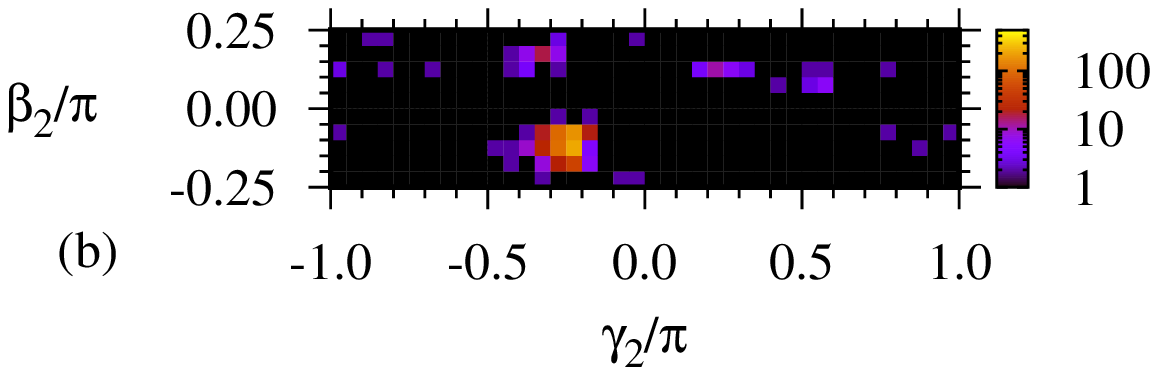}
    \includegraphics[width=9cm,height=8cm,keepaspectratio,angle=0,trim={0 2.5cm 0 2.5cm},clip]{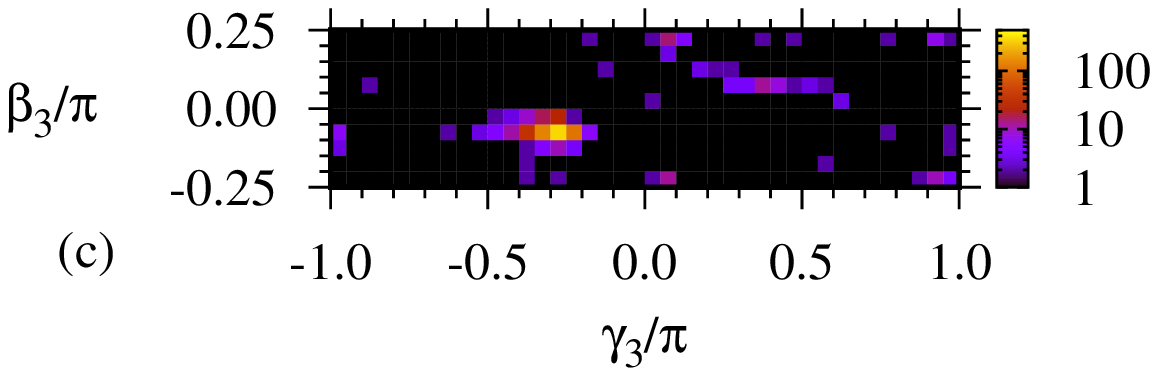}
    \caption{Angle patterns at depth $p=3$ for graphs with $n=7$ vertices, for the first layer parameters $(\gamma_1,\beta_1)$ in (a), $(\gamma_2,\beta_2)$ in (b), and $(\gamma_3,\beta_3)$ in (c).}
    \label{p=3 angle fig}
\end{figure}

Figure \ref{p=3 angle fig} shows the optimized angle distributions at $p=3$ and $n=7$; Table \ref{p=3 angle table} shows the fractions of graphs that deviate from $(\Gamma,\Beta)$.  These continue the pattern:  the majority of angles are concentrated in a single cluster in $(\Gamma,\Beta)$ at each layer, with additional small clusters of angles distributed unpredictably over the parameter space.  The fractions of graphs in these clusters increases with $p$.  Deviations from the angle patterns occur most often in the final layer of angles and in combinations connecting the final layer with earlier layers, with relatively large $ \mathcal{D}^{(3)}_{n,p} \approx \mathcal{D}^{(23)}_{n,p} \approx \mathcal{D}^{(123)}_{n,p}$.  Typically, smaller fractions of graphs deviate in the earlier layers only, as in $\mathcal{D}^{(1)}_{n,p}, \mathcal{D}^{(2)}_{n,p}$, and $\mathcal{D}^{(12)}_{n,p},$ or in disjoint layers, as in  $\mathcal{D}^{(13)}_{n,p}$.  The total fractions of graphs that deviate from the patterns $\mathcal{D}_{n,p}$ decreases with $n$ and increases with $p$.

\begin{table*}
    \centering
    \begin{tabular}{| c | c | c | c | c | c | c | c | c |  c |}
        \hline
        \ $n$ \ & \ $p$ \ & $\mathcal{D}^{(1)}_{n,p}$ & $\mathcal{D}^{(2)}_{n,p}$ & $\mathcal{D}^{(3)}_{n,p}$ & $\mathcal{D}^{(12)}_{n,p}$ & $\mathcal{D}^{(13)}_{n,p}$ & $\mathcal{D}^{(23)}_{n,p}$ & $\mathcal{D}^{(123)}_{n,p}$ & $\mathcal{D}_{n,p}$\\         
          \hline
         2 &  3 & 0.000 & 0.000 & 0.000 & 0.000 & 0.000 & 0.000 & 0.000 & 0.000\\
        \hline
        3 &  3 & 0.000 & 5.000$\times 10^{-1}$ & 0.000 & 0.000 & 0.000 & 0.000 & 0.000 & 5.000$\times 10^{-1}$ \\
        \hline
        4 & 3 & 0.000 & 0.000 & 0.000 & 1.667$\times 10^{-1}$ & 0.000 & 1.667$\times 10^{-1}$ & 1.667$\times 10^{-1}$ & 5.000$\times 10^{-1}$ \\
        \hline
        5 & 3 & 0.000 & 4.762$\times 10^{-2}$ & 9.524$\times 10^{-2}$ & 0.000 & 4.762$\times 10^{-2}$ & 2.381$\times 10^{-1}$ & 4.762$\times 10^{-1}$ & 4.762$\times 10^{-1}$  \\
        \hline
        6 & 3 & 8.929$\times 10^{-3}$ & 0.000 & 6.250$\times 10^{-2}$ &8.929$\times 10^{-3}$ & 1.786$\times 10^{-2}$ &  1.250$\times 10^{-1}$ & 8.036$\times 10^{-2}$ & 3.036$\times 10^{-1}$\\
        \hline
        7 & 3 & 0.000 & 1.172$\times 10^{-3}$ & 3.517$\times 10^{-2}$ & 3.517$\times 10^{-3}$ & 2.345$\times 10^{-3}$ & 5.627$\times 10^{-2}$ & 3.048 $\times 10^{-2}$ & 1.290$\times 10^{-1}$\\
        \hline
        8 & 3 & 0.000 & 4.498$\times 10^{-4}$ & 1.079$\times 10^{-2}$ & 5.397$\times 10^{-4}$ & 9.895$\times 10^{-4}$ & 1.430$\times 10^{-2}$ & 1.934$\times 10^{-2}$ & 4.642$\times 10^{-2}$\\
        \hline  
        9 & 3 & 0.000 & 6.511 $\times 10^{-5}$ & 2.731 $\times 10^{-3}$ & 4.979 $\times 10^{-5}$ & 1.264 $\times 10^{-4}$ & 2.214 $\times 10^{-3}$ & 6.764 $\times 10^{-3}$ & 1.195 $\times 10^{-2}$ \\
        \hline
    \end{tabular}
    \caption{Fractions of graphs with optimized angles that deviate from the region $(\Gamma,\Beta)$ at $p=3$, similar to Tables \ref{p=1 angle table}-\ref{p=2 angle table}.}
    \label{p=3 angle table}
\end{table*}

The deviations in the clusters of angles away from $(\Gamma,\Beta)$ appears to be due to differences in graph structure.  We have confirmed that optimal angles deviate from $(\Gamma,\Beta)$ for some graphs. While parameter optimization is limited by the sampling of random seeds with BFGS, these deviations are not attributed to limits on sampling. We expect the clusters contain graphs with similar structures, with more graphs in the clusters at higher $p$ indicating a greater sensitivity to graph structure features.  Angles that optimize a graph at a given $p$ are not necessarily close to the angles that optimize the same graph at $p+1$. However, for most graphs at the $n$ and $p$ tested here, the angle variations are minimal and contained in $(\Gamma,\Beta)$.  This motivates a heuristic approach to identifying optimized angles for most graphs, developed in the next section.

\subsection{Median Angles} \label{median angle section}

We consider how patterns of optimized angles may identify good approximate angles for most graphs and greatly reduce the computational cost of searching for angles with BFGS.  Similar uses of angle patterns have been considered for 3-regular graphs \cite{zhou2020quantum}, families of structured graphs \cite{brandao2018concentration}, and small samples of random graphs \cite{crooks2018performance}.
 
\begin{figure}
    \includegraphics[width=9cm,height=9cm,keepaspectratio,angle=0,trim={0cm 0cm 0cm 0cm}, clip]{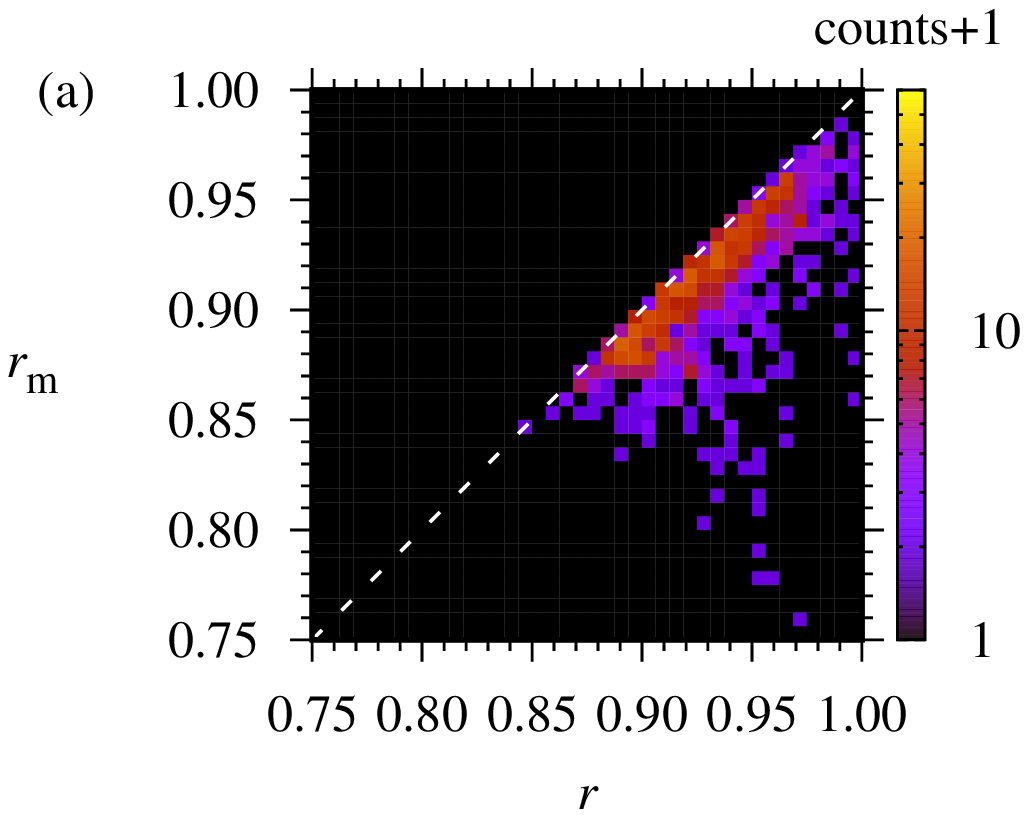}
    \includegraphics[width=9cm,height=9cm,keepaspectratio,angle=0,trim={0cm 0cm 0 0cm},clip]{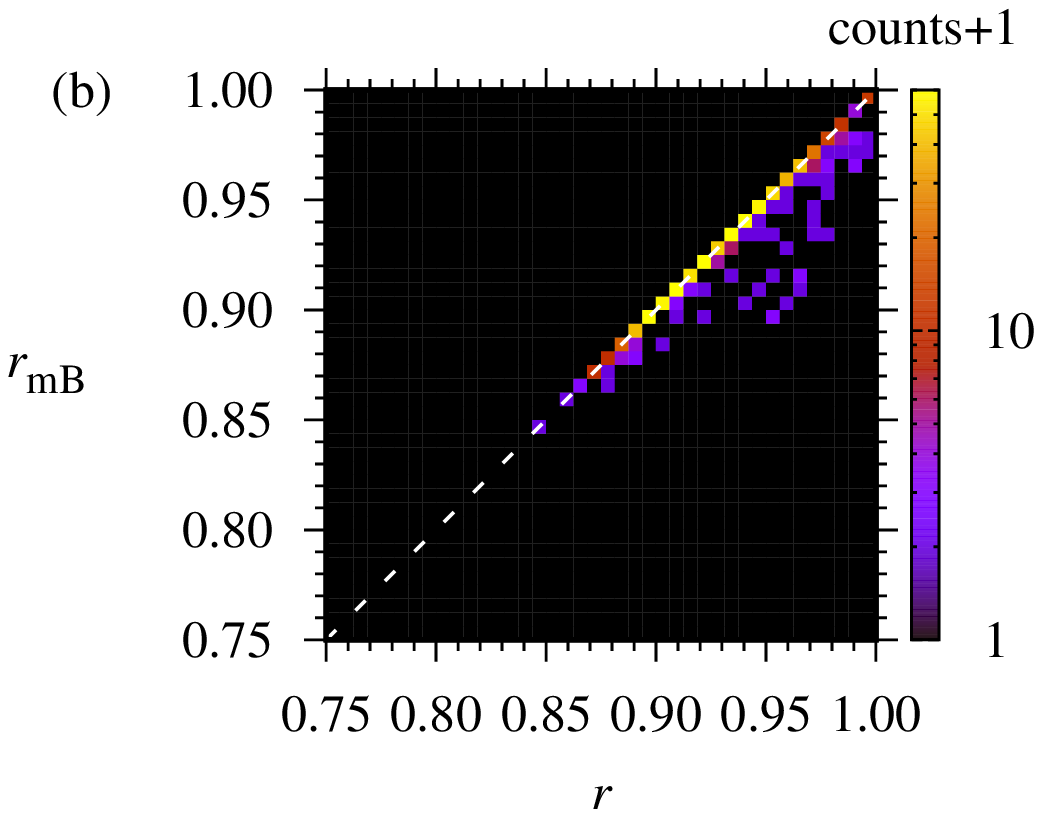}
    \caption{Two-dimensional histograms comparing the approximation ratio from the full BFGS search to results using the median angles without optimization (a) and using the the median angles as seeds in BFGS (b), for $n=7$ and $p=3$. Dashed white lines indicate the diagonals where $r_\mathrm{m} = r$ and $r_\mathrm{mB} = r$.}
    \label{median angle r fig}
\end{figure}

\begin{figure}
    \includegraphics[width=9cm,height=9cm,keepaspectratio,angle=0,trim={0cm 0cm 0cm 0cm},clip]{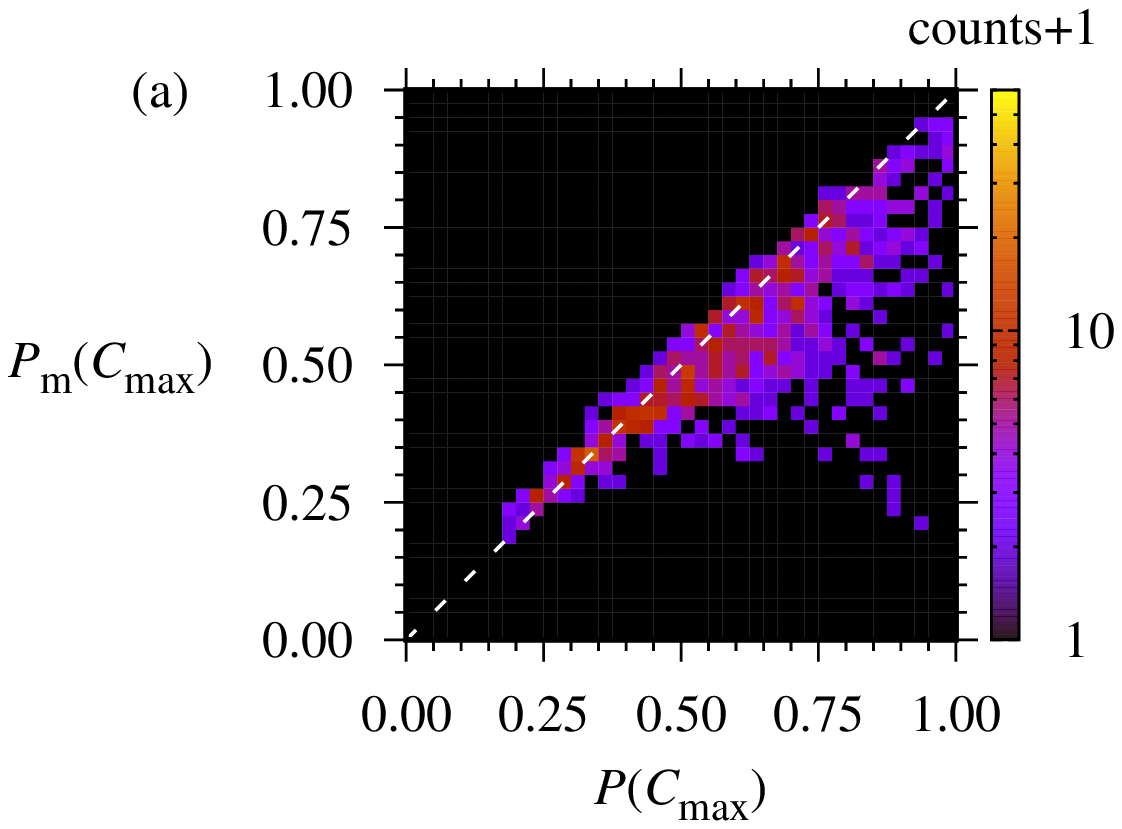}
    \includegraphics[width=9cm,height=9cm,keepaspectratio,angle=0,trim={0cm 0cm 0cm 0cm},clip]{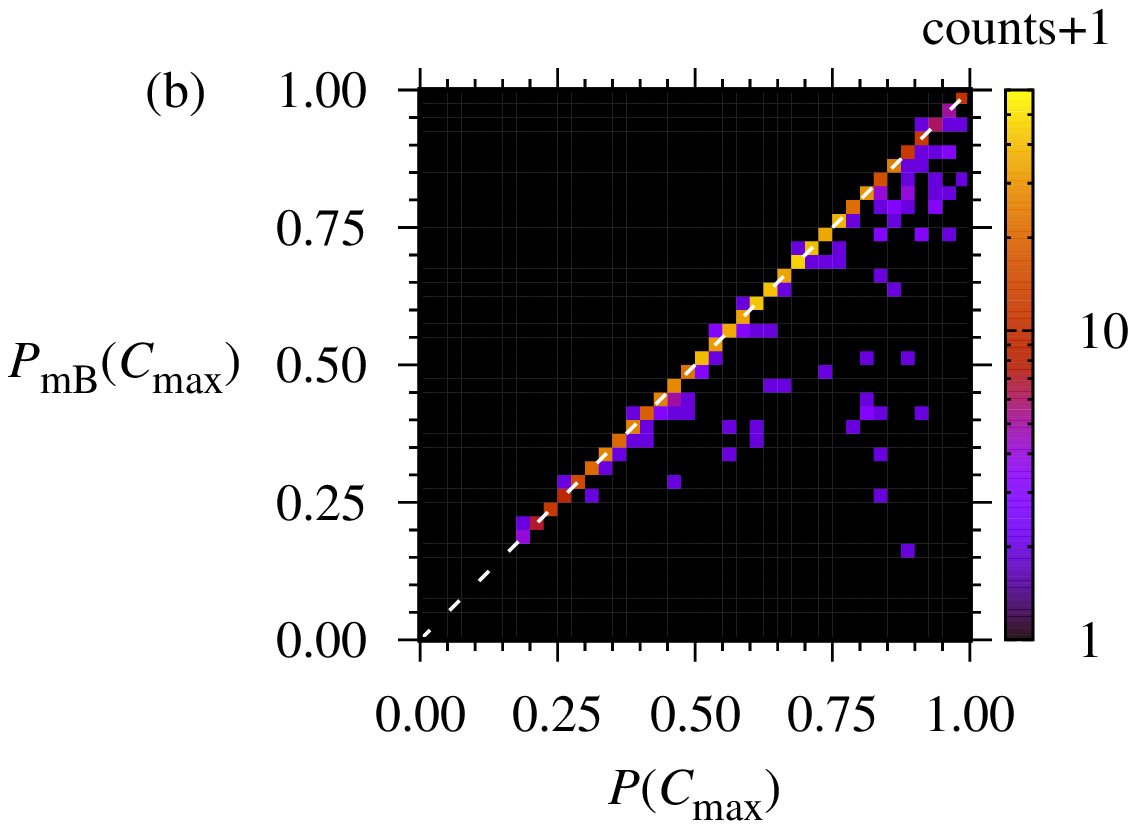}
    \caption{Histograms similar to Fig.~\ref{median angle r fig} but showing the probability of obtaining a maximum cut.}
    \label{median angle p fig}
\end{figure}

We define a set of angles that follows the angle patterns by first taking the median $\gamma_1$ over all $\gamma_1$ for graphs with $n=7$ and $p=3$, then define similar median $\gamma_j$ and $\beta_j$ for each $j$ to obtain the set of angles shown in Table \ref{median angles table}. We consider two approaches to QAOA that use these angles to avoid the computationally expensive random seeding of the standard BFGS approach. The first approach uses the median angles in the evolution of Eq.~(\ref{psi p}) without any optimization, results from this approach are denoted  $r_\mathrm{m}$ and $P_\mathrm{m}(C_\mathrm{max})$.  The second approach uses the median angles as seeds in a single BFGS optimization, results from this approach are denoted $r_\mathrm{mB}$ and $P_\mathrm{mB}(C_\mathrm{max})$. Figures \ref{median angle r fig}-\ref{median angle p fig} compare results from the median angle approaches at $n=7$ and $p=3$ to our previous results from BFGS optimization with hundreds of random seeds.  The results are visualized using two dimensional histograms on a logarithmic color scale.

\begin{table}
    \centering
    \begin{tabular}{| c | c | c | c | c | c | }
        \hline
        $\beta_1/\pi$ &  $\beta_2/\pi$ &  $\beta_3/\pi$ &  $\gamma_1/\pi$ &  $\gamma_2/\pi$ &  $\gamma_3/\pi$  \\
        \hline
         -0.15244 & -0.10299 & -0.06517 & -0.12641 & -0.24101 & -0.27459 \\
        \hline
    \end{tabular}
    \caption{Median optimized angles at $n=7$ and $p=3$, shown to five decimal places.}
    \label{median angles table}
\end{table}

Figure~\ref{median angle r fig}(a) compares the standard $r$ from the full BFGS search to the $r_\mathrm{m}$ from the median angles without any optimization. The results are concentrated near the diagonal, shown by the white dotted line, and the mean and standard deviation of the difference $r - r_\mathrm{m}$ is small, as seen in Table \ref{median angle difference table}.  The median angles $r_\mathrm{m}$ give a good approximation to the $r$ from the full BFGS search.

\begin{table*}
    \centering
    \begin{tabular}{| c | c | c | c | c | }
        \hline
        & $r - r_\mathrm{m}$ &  $r - r_\mathrm{mB}$ &  $P(C_\mathrm{max}) - P_\mathrm{m}(C_\mathrm{max}) $ &  $P(C_\mathrm{max}) - P_\mathrm{mB}(C_\mathrm{max})$ \\
        \hline
         mean &  2.2434 $\times 10^{-2}$ & 1.4198 $\times 10^{-3}$   &  7.2206 $\times 10^{-2}$  & 1.2641 $\times 10^{-2}$  \\
        \hline
         standard deviation & 2.4664 $\times 10^{-2}$ & 6.2716 $\times 10^{-3}$ & 1.0585 $\times 10^{-1}$ & 5.9882 $\times 10^{-2}$ \\
        \hline
    \end{tabular}
    \caption{Mean and standard deviations of the differences between quantities $r_\mathrm{m}, r_\mathrm{mB},P_\mathrm{m}(C_\mathrm{max})$, and $P_\mathrm{mB}(C_\mathrm{max}) $ calculated with the median angles and quantities $r$ and $P(C_\mathrm{max}) $ calculated from BFGS optimizations with random seeds, see text for details. }
    \label{median angle difference table}
\end{table*}

 Figure \ref{median angle r fig}(b) compares the standard $r$ against the $r_{\mathrm{mB}}$ from single BFGS optimizations using the median angles as seeds.  The $r_{\mathrm{mB}}$ are significantly improved in comparison with the $r_\mathrm{m}$, with 87\% of the results satisfying  $r_\mathrm{mB} = r$ up to an additive factor of $10^{-6}$, with significantly reduced mean and standard deviation of the difference $r_\mathrm{mB} - r$ in Table \ref{median angle difference table}. The approximation ratio from the single runs of BFGS with the median angle seeds are typically identical or close to results from the BFGS search with random seeds and are calculated at a small fraction of the computational expense.

Figure \ref{median angle p fig} assesses the probability of obtaining the maximum cut in the median angle approaches. The $P_\mathrm{m}(C_\mathrm{max})$ in Fig.~\ref{median angle p fig}(a) roughly follow the $P(C_\mathrm{max})$ along the diagonal but with some considerable spread below the diagonal; note the difference in scale in comparison with the $r$ in Fig.~\ref{median angle r fig}.  There is also some spread above the diagonal, indicating the median angles give higher probabilities for the maximum cut for some graphs.  The mean and standard deviation of the difference $P_\mathrm{m}(C_\mathrm{max}) - P(C_\mathrm{max})$ is shown in Table \ref{median angle difference table}; they are small but larger than the corresponding values for $r-r_\mathrm{m}$.  Overall, the median angles give probabilities for the maximum cut that are typically close to the full BFGS results, but the probabilities are more sensitive to graph structures and vary significantly for some graphs.

Figure \ref{median angle p fig}(b) shows $P_\mathrm{mB}(C_\mathrm{max})$, where the median angles have been used in a single BFGS optimization.  The results are much closer to the diagonal than in Fig.~\ref{median angle p fig}(a) and the mean and standard deviation of the difference $P(C_\mathrm{max}) - P_\mathrm{mB}(C_\mathrm{max})$ are smaller than the corresponding quantities with $P_\mathrm{m}(C_\mathrm{max})$.  In comparison with the $r_\mathrm{mB}$, we again see the probability of obtaining the maximum cut is more sensitive to the choice of angles, with greater deviations for some graphs in the figures and a larger mean and standard deviation of the difference in Table \ref{median angle difference table}.  However, for most graphs the median angles work very well, with 87\% of graphs obtaining identical probabilities for the maximum cut using the median angles as seeds in BFGS and using the much more computationally expensive search over random seeds. 

\section{Conclusions} \label{conclusion}

We have presented results that identify empirical performance bounds on optimized instances of QAOA for MaxCut. Using numerical simulations, we investigated an exhaustive set of MaxCut instances for depths $p \leq 3$ on graphs with $n\leq 9$ vertices. We calculated optimal solutions using exact numerical search and brute force search at $p=1$ and $2$ and validated results from BFGS at depths $p \leq 3$.  The catalog of graph instances, optimized angles, and simulated states are available online \cite{dataset}.  

Our analysis used the approximation ratio $r$ and the probability for obtaining a maximum cut $P(C_\mathrm{max})$ as measures of QAOA performance. We observed that $r$ becomes more similar across graph structures as $p$ and $n$ increase and the Gaussian distributions of $r$ narrow. Most graphs at $n \leq 9$ exceed the Goemans-Williamson bound by $p=3$ in these narrow distributions, indicating viability of modest-depth QAOA to outperform GW. 

In contrast to the narrowing distributions of $r$, distributions of $P(C_\mathrm{max})$ were found to broaden with increasing $p$. We attributed the difference to design of the cost function and its corresponding spectrum. A preponderance of nearly-optimal states skews the optimization of $r$ away from the much smaller set of truly optimal states. While this yields a large, uniform value for $r$, it leaves $P(C_\mathrm{max})$ less constrained. One alternative is to focus optimization on state preparation to ensure the largest values of $P(C_\mathrm{max})$, for example by choosing gate angles that optimize a non-linear function such as $\langle \exp(\alpha\hat C)\rangle$ \cite{LiLi2020Gibbs}. For $p>0$, we observed an exponential decay in $P(C_\mathrm{max})$ with respect to $n$. The rate constant, $k_p$, was found to decrease as $p$ increases, and this raises the question as to whether such exponential behavior can predict performance metrics at larger $n$ and $p$.

The patterns observed in the optimized angles across this exhaustive set of instances mirrors previous results for narrower cases \cite{brandao2018concentration,Bartschi2020MaxkCover,zhou2020quantum,crooks2018performance,Shaydulin2020CaseStudy}.  Using these patterns as a search heuristic demonstrated high quality results for most graphs and required a significantly smaller computational cost than BFGS search with random seeding. Identifying the prevalence of these patterns at larger $n$ and $p$ as well as the correlation with graph properties can enable new search heuristics, for example, for  general QAOA-like algorithms, as suggested by the work of Cook \etal \cite{Bartschi2020MaxkCover}.
  
\acknowledgments

P.~C.~L.~thanks Zak Webb and Yan Wang for discussing time-reversal symmetry. This work was supported by DARPA ONISQ program under award W911NF-20-2-0051. J. Ostrowski acknowledges the Air Force Office of Scientific Research award, AF-FA9550-19-1-0147. G.\  Siopsis  acknowledges the Army Research Office award W911NF-19-1-0397. J.\ Ostrowski and G.\ Siopsis acknowledge the National Science Foundation award OMA-1937008. This research used resources of the Compute and Data Environment for Science (CADES) at the Oak Ridge National Laboratory, which is supported by the Office of Science of the U.S. Department of Energy under Contract No. DE-AC05-00OR22725.

\appendix

\section*{Appendix: Angle symmetries}

In this Appendix we discuss details of the angle symmetries from Section \ref{angle symmetry section}.  Most of the symmetries have been described previously by Zhou {\it et.~al.}~\cite{zhou2020quantum}, although we note two of the symmetries they described for regular graphs apply more broadly to graphs where every vertex has even degree or every vertex has odd degree.  Each symmetry relates different sets of angles $(\bm \gamma, \bm \beta)$ and $(\bm \gamma', \bm \beta')$ that give the same approximation ratio in Eq.~(\ref{expectation c eq}), 
\begin{equation} \label{symmetry} \langle \hat C(\bm \gamma,\bm\beta) \rangle = \langle \hat C(\bm\gamma', \bm \beta')\rangle,  \end{equation}

\noindent where 
$$\bm \gamma = (\gamma_1,...,\gamma_{q-1},\gamma_q,\gamma_{q+1},...,\gamma_p)$$
\begin{equation} \label{generic angles} \bm \beta = (\beta_1,...,\beta_{q-1},\beta_q,\beta_{q+1},...,\beta_p), \end{equation}

\noindent with $(\bm \gamma', \bm \beta')$ related to $(\bm \gamma, \bm \beta)$ in different ways for the different symmetries.  An example is  periodic behavior of the $\beta_q$ angles over intervals of $\pi/2$, we use this to restrict every $\beta_q$ component to the interval $-\pi/4 \leq \beta_q \leq \pi/4$, as described by Zhou {\it et.~al.}  There are a variety of additional symmetries we describe below, with derivations in subsequent subsections.

The first type of symmetry applies to graphs where every vertex has even degree.  In this case the $\gamma_q$ angles are periodic over intervals of $\pi$ since all the eigenvalues of $\hat C$ are even, as shown in detail in the next subsection.   Thus the symmetry of Eq.~(\ref{symmetry}) holds for any pair of angles with $\bm \beta = \bm \beta'$ and
\begin{equation} \label{even symmetry} \bm \gamma' = (\gamma_1,...,\gamma_{q-1},\gamma_q \pm \pi,\gamma_{q+1},...,\gamma_p), \end{equation}

\noindent where $\bm \gamma'$ differs from $\bm \gamma$ by a shift $\gamma_q' = \gamma_q \pm \pi$ for any $q$.  We use this symmetry to organize the angle distributions so $|\gamma_q| \leq \pi/2$ for all $q$ for graphs where every vertex degree is even.  We search for generic optimized angles in our implementation of the BFGS algorithm from Section \ref{BFGS section}, but if we find a $|\gamma_q| > \pi/2$ then we add or subtract $\pi$ to get a $|\gamma_q'| \leq \pi/2$. 

The second type of symmetry applies to graphs where every vertex has odd degree.  This gives a joint symmetry in both sets of angles in Eq.~(\ref{symmetry}),
$$\bm \gamma' = (\gamma_1,...,\gamma_{q-1},\gamma_q \pm \pi,\gamma_{q+1},...,\gamma_p)$$
\begin{equation}\label{odd symmetry} \bm \beta' = (\beta_1,...,\beta_{q-1},-\beta_q,-\beta_{q+1},...,-\beta_p).\end{equation}

\noindent In Eq.~(\ref{odd symmetry}), $\bm\gamma'$ differs from $\bm \gamma$ in the $q$th component, $\gamma_q' = \gamma_q \pm \pi$ for any $q$, and $\bm \beta'$ differs from $\bm \beta$ in the sign of all subsequent components, $\beta_r'= -\beta_r$ for all $r \geq q$.  The proof follows from Pauli operator commutation relations applied to the two unitary operators of Eqs.~(\ref{UC})-(\ref{UB}), as shown in detail later.     In our calculations, we use this symmetry to organize the angles so $|\gamma_q| \leq \pi/2$ for all $q$ for  graphs where every vertex degree is odd, similar to how we organize the angles when all the vertex degrees are even. 

The third analytic symmetry we use is the ``time-reversal" symmetry \cite{zhou2020quantum}
\begin{equation} \label{time reversal} (\bm \gamma',\bm\beta')  = (-\bm\gamma, -\bm \beta).  \end{equation}

\noindent The symmetry is related to the $\hat B$ and $\hat C$ operators in Eqs.~(\ref{UC})-(\ref{UB}) and the initial state $\vert \psi_0\rangle$ of Eq.~(\ref{initial state}), which  have real-valued matrix elements and coefficients in the computational basis.  We give a proof at the end of the Appendix.  We use this symmetry to always transform to angles with $\beta_1 \leq 0$.  The $\beta_q$ for $q > 1$ can be positive or negative.

We find additional symmetries in the optimized angles for a small subset of graphs. For these, we report the angles with the most component-pairs $(\gamma_i,\beta_i) \in (\Gamma, \Beta)$ from Eq.~(\ref{GammaBeta}).  Using the angles with the most components in $(\Gamma, \Beta)$ is designed to emphasize the angle patterns in Section \ref{angle section}.  The number of graphs $M(p)$ we used the $(\Gamma, \Beta)$ symmetry on is shown for each $n$ and $p$ in Table \ref{symmetry table}.  The symmetry was used most often when a graph became fully optimized.  For example, we found two sets of angles for an $n=3$ graph that optimized the cost function $\langle \hat C \rangle = C_\mathrm{max}$; we saved the angles with the most components in $(\Gamma,\Beta)$.  Asterisks in the table denote each graph which we used the rule on for that $n$ and $p$ was fully optimized.  Note when graphs are optimized at some $p$ we do not simulate them at depths $p' > p$, so they do not carry over to columns for greater depths in the table.  Overall, the rule is applied to a very limited subset of the graphs we study.

\begin{table}
    \centering
    \begin{tabular}{| c | c | c | c |}
        \hline
        \ $n$ \ & $M(1)$ & $M(2)$ & $M(3)$\\
        \hline
         2 & 0 & -- & --  \\
        \hline
        3 & 0 & 1* & --\\
        \hline
         4 & 0 & 2* & 0 \\
        \hline
        5 & 0 & 2* & 1* \\
        \hline
        6 & 0 & 1 & 3*\\
        \hline
        7 & 0 & 0 & 2*\\
        \hline
        8 & 0 & 1 & 1\\
        \hline
        9 & 0 & 0 & 1*\\
        \hline
    \end{tabular}
    \caption{Numbers of graphs $M(p)$ where we found degeneracies and used the rule of saving the angles with the most components in $(\Gamma,\Beta)$ following Eq.~(\ref{GammaBeta}), for various depths $p$.  Numbers with asterisks denote all the graphs we used the rule on were fully optimized, dashes denote all the graphs were optimized at smaller depths.  }
    \label{symmetry table}
\end{table}

\subsection{Symmetry when all vertices have even degree} \label{even symmetry section}

When all the vertices have even degree, the angle components $\gamma_q$ are periodic over intervals of $\pi$, as discussed around Eq.~(\ref{even symmetry}).  To demonstrate equivalence of $\langle \hat C \rangle$ in Eq.~(\ref{symmetry}) for the angles $\bm \gamma'$ from Eq.~(\ref{even symmetry}) and $\bm \gamma$ from Eq.~(\ref{generic angles}), in the next paragraph we show $\hat C$ of Eq.~(\ref{cost operator}) has only even eigenvalues when each vertex has even degree.  Then $C(z) = 2m_z$ for all $z$, where $m_z$ is an integer.  The periodicity over $\pi$ follows since the matrix elements from the unitary operator $\hat U(\hat C,\gamma_q)$ of Eq.~(\ref{UC}) are invariant under changes $\gamma_q \to \gamma_q \pm \pi$, since
$$ \langle z \vert\hat U(\hat C,\gamma_q)\vert z \rangle  = \exp(\mp i2m_z\pi)\exp(-i 2m_z \gamma_q) $$ \begin{equation}\label{even symmetry relation} =  \langle z \vert\hat U(\hat C,\gamma_q \pm \pi)\vert z \rangle, \end{equation}

\noindent where the factor $1 = \exp(\mp i2m_z\pi)$ connects the two expressions.

To show the eigenvalues $C(z)$ are all even when all the vertex degrees are even, we begin by showing  there exists a single bitstring $z^{(0)}$ for which $C(z^{(0)})$ is even. Then we show that if $C(z)$ is even for any $z$ and all the vertex degrees are even, then modifying any bit $z_k$ in the bitstring $z$ to get a new bitstring $z'$ will give a $C(z')$ that is also even.  Together these imply that every bitstring has even $C(z)$ when all the vertex degrees are even, since any bitstring can be made from a series of modifications to $z^{(0)}$ and every modification gives an even $C(z)$.

First consider the zero bitstring $z^{(0)} = (0,0,...,0)$. From Eqs.~(\ref{cost function})-(\ref{edge cost function}) this has $C(z^{(0)}) = 0$ which is even.  Next consider an arbitrary bitstring $z  = (z_{n-1}, z_{n-2}, ..., z_k, ..., z_0)$ for which $C(z)$ is even.  Suppose we flip a bit $z_k$ to make a new bitstring $z'  = (z_{n-1}, z_{n-2}, ..., z_k', ..., z_0)$ where $z_k' \neq z_k$ but the $z_{j \neq k}$ are the same.  The change $z_k \to z_k'$ will change the value of each $C_{\langle j,k\rangle}$ from Eq.~(\ref{edge cost function}) that is associated with $z_k$, so that if $C_{\langle j,k\rangle}(z)$ = 1 then $C_{\langle j,k\rangle}(z') = 0$ and vice-versa. The total number of edge terms $C_{\langle j,k\rangle}$ can be separated into $\kappa$ terms that increase in value and $\delta$ terms that decrease in value when $z \to z'$. The value of the cost function for $z'$ can then be expressed as $C(z') = C(z) + \kappa - \delta$.  The number of edge components $C_{\langle j,k\rangle}$ that depend on $z_k$ is even when each vertex degree is even, so  $\delta + \kappa = 2m$ for some integer $m$.  This implies that $\kappa$ and $\delta$ are either both even or both odd; either way, the difference $\kappa-\delta$ is even. Since $C(z)$ is even by assumption, $C(z') = C(z) + \kappa - \delta$ is also even.  

We have shown there is a bitstring $z^{(0)}$ for which $C(z^{(0)})$ is even and we have shown changing any such bitstring gives a new $C(z')$ which is also even.  This implies $C(z)$ is even for all $z$ for every graph where each vertex degree is even.  By Eq.~(\ref{even symmetry relation}) the angle components $\gamma_q$ are periodic over intervals of $\pi$ for these graphs.

\subsection{Symmetry when all vertices have odd degree}

When all the vertices have odd degree, there is a symmetry Eq.~(\ref{symmetry}) for angle-pairs $(\bm \gamma, \bm \beta)$ and $(\bm \gamma',\bm\beta')$ from Eqs.~(\ref{generic angles}) and (\ref{odd symmetry}) respectively.   Only graphs with an even number of vertices can have this symmetry since it is impossible to have a graph with an odd number of vertices where every vertex has odd degree. We prove the symmetry holds in a simple case with $p=1$, the extension to $p>1$ uses similar reasoning.  We begin by considering the relations between the unitary operators with $(\bm \gamma, \bm \beta)$ and $(\bm \gamma',\bm\beta')$, then use the analysis to relate the time evolution and probabilities $P(z)$ for states with the different angles.

Let $\gamma_1' = \gamma_1 \pm \pi$. The unitary operator $\hat U(\hat C, \gamma_1)$ from Eq.~(\ref{UC}) is a product of unitary-operator components for each edge, a single component can be expressed as
$$  \exp(-i \gamma_1 \hat C_{\langle j,k\rangle}) = \exp\left(-i \frac{\gamma_1}{2}(\hat{\mathbb{1}} - \hat Z_j \hat Z_k)\right)  $$ \begin{equation} \label{Uc1} =\exp\left(-i \frac{\gamma_1' \mp \pi}{2}\right)\exp\left(i \frac{\gamma_1' \mp \pi}{2}\hat Z_j \hat Z_k\right). \end{equation}

\noindent Separate out a term $\exp\left(\mp i (\pi/2)\hat Z_j \hat Z_k\right)= \mp i \hat Z_j \hat Z_k$ to obtain
\begin{equation}\label{Uc2} \exp(-i \gamma_1 \hat C_{\langle j,k\rangle}) = e^{i\delta} \hat Z_k \hat Z_j  \exp(-i \gamma_1' \hat C_{\langle j,k\rangle}),  \end{equation}

\noindent where $e^{i\delta}$ is an overall phase.  

Now consider the unitary operator $\hat U(\hat C, \gamma_1)$ of Eq.~(\ref{UC}).  Using Eq.~(\ref{Uc2}) we have
$$ \hat U(\hat C, \gamma_1) = \prod_{\langle j,k\rangle} \exp(-i \gamma_1 \hat C_{\langle j,k\rangle}) $$
\begin{equation} \label{Uc3} = e^{i\eta}\left(\prod_{j=0}^{n-1} \hat Z_j\right) \hat U(\hat C, \gamma_1'), \end{equation}

\noindent where $e^{i\eta}$ is a phase factor.  The term $\prod_{j=0}^{n-1} \hat Z_j$ comes from the product of $\hat Z_j \hat Z_k $ for all the edges---each $\hat Z_j$ is raised to an odd power in the product since each vertex degree is odd and using $(\hat Z_j)^2 = \hat{\mathbb{1}}$ reduces this to $\prod_{j=0}^{n-1} \hat Z_j $.

We will use Eq.~(\ref{Uc3}) to simplify the time evolution of $\vert \psi_p(\gamma_1,\beta_1)\rangle$ in Eq.~(\ref{psi p}).  This includes the unitary operator $\hat U(\hat B, \beta_1)$ from Eq.~(\ref{UB}), which shows up in the product  $\hat U(\hat B, \beta_1) \hat U(\hat C, \gamma_1)$.  Our next goal will be to use commutation relations to move the $\prod_{j=0}^{n-1} \hat Z_j$ to the left side of the product.

Express the $\hat U(\hat B,\beta_1)$ from Eq.~(\ref{UB}) as
\begin{equation} \hat U(\hat B, \beta_1) = \prod_{j=0}^{n-1}\left(\cos(\beta_1)\hat{\mathbb{1}} - i \sin(\beta_1)\hat X_j\right). \end{equation}

\noindent Now consider the product of the unitary operators
$$ \hat U(\hat B, \beta_1) \hat U(\hat C, \gamma_1)$$  \begin{equation} = e^{i\eta}\prod_{j=0}^{n-1}\left(\cos(\beta_1)\hat{\mathbb{1}} - i \sin(\beta_1)\hat X_j\right)\left(\prod_{j=0}^{n-1} \hat Z_j \right)\hat U(\hat C, \gamma_1') \end{equation}

\noindent The Pauli operators anticommute so 
\begin{equation} \label{Uc4} \hat U(\hat B, \beta_1) \hat U(\hat C, \gamma_1) = e^{i\eta}\left(\prod_{j=0}^{n-1} \hat Z_j \right)\hat U(\hat B, -\beta_1) \hat U(\hat C, \gamma_1'). \end{equation} 

We are now ready to show the angles $(\gamma_1, \beta_1)$ and $(\gamma_1',\beta_1')$ give equivalent $\langle \hat C \rangle$ in Eq.~(\ref{expectation c eq}), where $\gamma_1' = \gamma_1 \pm \pi$ and $\beta_1' = - \beta_1$. To demonstrate the equivalence we show the computational basis state probabilities $P(z)$ are the same for both $(\gamma_1, \beta_1)$ and $(\gamma_1',\beta_1')$. 

The probability amplitude for a basis state $\vert z \rangle$ is
\begin{equation} \langle z \vert \psi (\gamma_1,\beta_1)\rangle = \langle z \vert \hat U(\hat B, \beta_1) \hat U(\hat C,\gamma_1) \vert \psi_0\rangle  \end{equation}

\noindent Using Eq.~(\ref{Uc4}) this can be expressed as 
$$ \langle z \vert \psi(\gamma_1,\beta_1) \rangle =\langle z \vert e^{i\eta}\left(\prod_{j=0}^{n-1} \hat Z_j \right)\hat U(\hat B, \beta_1') \hat U(\hat C, \gamma_1') \vert \psi_0\rangle $$
\begin{equation} = e^{i\eta}\langle z \vert \left(\prod_{j=0}^{n-1}\hat Z_j \right)\vert \psi(\gamma_1',\beta_1')\rangle \end{equation}

\noindent Squaring the amplitudes we obtain
$$ \vert\langle z \vert \psi(\gamma_1,\beta_1) \rangle|^2 = $$
\begin{equation} \langle \psi(\gamma_1',\beta_1') \vert \left(\prod_{j=0}^{n-1} {\hat Z_j}^{\dag} \right)\vert z \rangle\langle z \vert \left(\prod_{j=0}^{n-1} {\hat Z_j} \right)\vert \psi(\gamma_1',\beta_1')\rangle \end{equation}

\noindent The $\left(\prod_{j=0}^{n-1} {\hat Z_j} \right)$ can be moved to the left since it commutes with $\vert z \rangle\langle z \vert$, then using $\left(\prod_{j=0}^{n-1} {\hat Z_j}^\dag \right)\left(\prod_{j=0}^{n-1} {\hat Z_j} \right) = \hat{\mathbb{1}}$ gives
\begin{equation} P(z) = \vert\langle z \vert \psi(\gamma_1,\beta_1) \rangle|^2 = \vert\langle z \vert \psi(\gamma_1',\beta_1') \rangle|^2, \end{equation}

\noindent where $P(z)$ is the probability of $\vert z \rangle$ from Eq.~(\ref{pz}).  The $P(z)$ are the same for $(\gamma_1, \beta_1)$ and $(\gamma_1',\beta_1')$ so $\langle \hat C(\gamma_1,\beta_1)\rangle = \langle \hat C(\gamma_1',\beta_1')\rangle$ in Eq.~(\ref{expectation c eq}), which is the desired symmetry.

\subsection{Time-reversal symmetry}

We finally consider the ``time-reversal" symmetry with the angles of Eqs.~(\ref{generic angles}) and (\ref{time reversal}) in the symmetry relation Eq.~(\ref{symmetry})  \cite{zhou2020quantum}. The symmetry is related to structures of the $\hat B$ and $\hat C$ operators, which have real-valued matrix elements in the computational basis in Eqs.~(\ref{mixer}) and (\ref{cost operator}), and the structure of the initial state $\vert \psi_0\rangle$, which has real-valued coefficients in the computational basis in Eq.~(\ref{initial state}).  To demonstrate the symmetry we calculate generic computational basis probabilities $P(z)$ for states with both sets of angles and show they are equal, thus the $\langle \hat C \rangle$ are equal following Eq.~(\ref{expectation c eq}).  

Consider the probability amplitude $\langle z \vert \psi_p(\bm \gamma,\bm\beta)\rangle$ for a single basis state $\vert z \rangle$.  From Eqs.~(\ref{initial state})-(\ref{UB}) this is 
\begin{equation} \label{prob amp} \langle z \vert \psi_p(\bm \gamma,\bm\beta)\rangle = \frac{1}{\sqrt{2^n}} \sum_{z'} \langle z \vert \left(\prod_{q=1}^p  e^{-i\beta_q\hat B} e^{-i\gamma_q\hat C} \right) \vert z'\rangle. \end{equation}
\noindent Taking the complex conjugate of Eq.~(\ref{prob amp}) only changes the sign of the angle terms since the $\hat B$ and $\hat C$ matrices have real-valued matrix elements in the computational basis, thus
\begin{equation} \langle z \vert \psi_p(\bm \gamma,\bm\beta)\rangle^* =  \langle z \vert \psi_p(-\bm \gamma,-\bm\beta)\rangle. \end{equation}
\noindent The Born probabilities $P(z)$ from Eq.~(\ref{pz}) are the same for both states since
$$P(z) = \vert \langle z \vert \psi_p(\bm \gamma,\bm\beta)\rangle \vert^2 = \vert \langle z \vert \psi_p(\bm \gamma,\bm\beta)\rangle^*\vert^2$$ \begin{equation}= \vert \langle z \vert \psi_p(-\bm \gamma,-\bm\beta)\rangle \vert^2, \end{equation}

\noindent so the $\langle \hat C \rangle$ are the same in Eq.~(\ref{expectation c eq}).

\bibliographystyle{unsrt}
\bibliography{references}

\end{document}